\documentclass[12pt,a4paper]{article}
\usepackage{epsfig}
\usepackage{epstopdf}
\usepackage{amsmath,amsfonts,amssymb}
\usepackage{cite}
\usepackage{colortbl}
\usepackage{pifont}

\providecommand{\openone}{\leavevmode\hbox{\small1\kern-3.8pt\normalsize1}}

\newcommand{\gm}{\gamma^\mu}
\newcommand{\Wm}{W_{\mu}}
\newcommand{\Zm}{Z_{\mu}}

\newcommand{\slx}{s_L^t}
\newcommand{\slu}{s_L^t}
\newcommand{\sld}{s_L^b}
\newcommand{\clx}{c_L^t}
\newcommand{\clu}{c_L^t}
\newcommand{\cld}{c_L^b}
\newcommand{\srx}{s_R^t}
\newcommand{\sru}{s_R^t}
\newcommand{\srd}{s_R^b}
\newcommand{\crx}{c_R^t}
\newcommand{\cru}{c_R^t}
\newcommand{\crd}{c_R^b}

\newcommand{\sqt}{\sqrt{2}}

\begin{document}

\begin{center}
	\begin{Large}
		{Interpreting the $W$-mass anomaly in the vectorlike quark models}
	\end{Large}
	
	\vspace{0.5cm}
	Junjie Cao, Lei Meng, Liangliang Shang, Shiyu Wang, Bingfang Yang \\[1mm]
	\begin{small}
		{\it School of Physics, Henan Normal University, Xinxiang 453007, China}
	\end{small}
\end{center}

\begin{abstract}
The new measurement of $W$-boson mass by the CDF collaboration revealed a remarkable $7\sigma$ disagreement with the Standard Model (SM) prediction. If confirmed by other experiments, then the disagreement strongly indicates the existence of new physics beyond the SM. In this work, seven vectorlike quark (VLQ) extensions of the SM are investigated to interpret the anomaly, and it is found that three can explain the anomaly in broad parameter space. The explanations are consistent with the constraints from oblique parameters, the LHC search for VLQs, the measurements of the properties for the top quark, bottom quark, and Higgs boson, and the perturbativity criterion. The typical size of the involved Yukawa coupling is around 1, which is comparable to the top quark Yukawa coupling in the SM. The other extensions, however, either predict a negative correction to the mass in reasonable parameter space or explain the anomaly by unnatural theoretical input parameters.
\end{abstract}

\section{INTRODUCTION}

Recently, the CDF collaboration at Fermilab reported their measured $W$-boson mass, $m_W^{CDF} = 80.4335 \pm 0.0094$~GeV\cite{CDF:2022hxs}, which deviates from the Standard Model (SM) prediction $m_W^{SM} = 80.357 \pm 0.006$~GeV\cite{ParticleDataGroup:2020ssz} by more than $7\sigma$. Even if all known and unknown theoretically higher-order corrections are included in the uncertainty estimation~\cite{Isaacson:2022rts}, there is still a discrepancy of about $5\sigma$. Such a large discrepancy, if confirmed by other measurements, strongly indicates the existence of new physics beyond the SM (BSM). Thus, it has attracted
considerable research attention ~\cite{Zhu:2022tpr,Kamleh:2021srn,Fan:2022dck,Lu:2022bgw,Athron:2022qpo,Yuan:2022cpw,Strumia:2022qkt,Yang:2022gvz,deBlas:2022hdk,Du:2022pbp,Tang:2022pxh,Cacciapaglia:2022xih,Sakurai:2022hwh,Fan:2022yly,Liu:2022jdq,Lee:2022nqz,Song:2022xts,Bagnaschi:2022whn,Paul:2022dds,Bahl:2022xzi,Asadi:2022xiy,DiLuzio:2022xns,Athron:2022isz,Gu:2022htv,Heckman:2022the,Babu:2022pdn,Balkin:2022glu,Biekotter:2022abc,Endo:2022kiw,Crivellin:2022fdf,Cheung:2022zsb,Du:2022brr,Heo:2022dey,Zheng:2022irz,Han:2022juu,Ahn:2022xeq,FileviezPerez:2022lxp,Ghoshal:2022vzo,Kawamura:2022uft,Ghorbani:2022vtv,Popov:2022ldh,Carpenter:2022oyg,Du:2022fqv,Bhaskar:2022vgk,Chowdhury:2022moc,Arcadi:2022dmt,Cirigliano:2022qdm,Borah:2022obi,He:2022zjz}.

Among the solutions to the discrepancy, extending the SM with vectorlike quarks (VLQs) is one of the most economic theoretical frameworks. Unlike the chiral quarks in the SM, the left- and right-handed components of the VLQs have the same transformation properties under the SM gauge group. As the VLQs are non-chiral, their mass terms do not arise from the Yukawa coupling to the SM Higgs field, which avoids the tight constraints on heavy fourth-generation quarks set by the Higgs boson data of the Large Hadron Collider (LHC)~\cite{ATLAS:2019nkf,CMS:2018gwt}. In addition, the vectorlike top quark partner among the VLQs was usually designed to cancel the largest quadratic divergence in the Higgs mass, which is caused by the top quark loop, so the fine-tuning problem of the SM can be alleviated (see, e.g., Refs.~\cite{Arkani-Hamed:2001nha,Arkani-Hamed:2002iiv,Berger:2012ec}). Correspondingly, VLQs have been widely adopted in building more complete models, such as the little Higgs~\cite{Arkani-Hamed:2002ikv,Schmaltz:2005ky} and composite Higgs models~\cite{Agashe:2004rs,Contino:2006qr}, to break the electroweak symmetry naturally.

In the VLQ extensions, the VLQs may mix with the SM quarks to form mass eigenstates and thereby modify their couplings to the $Z$, $W$, and Higgs bosons. Confronted with the atomic parity violation experiments, the electroweak precision observables (EWPOs) extracted from the large electron positron experiments, and the measurements of various low-energy flavor-conserving and flavor-violating processes, the VLQ mixings with the first two generations of quarks have been tightly limited, and only mixings with third-generation quarks may be sizable~\cite{Aguilar-Saavedra:2002phh}.  Fortunately, such a mixing pattern is favored from the theoretical perspective: The large Yukawa coupling of the top quark suggests a possible close connection of the top quark (and the left-handed component of the bottom quark due to the weak isospin symmetry) with any new physics related to the symmetry breaking and/or to the fermion mass hierarchy, which distinguishes the third-generation quarks from other quarks in the SM. It is emphasized that regardless of whether the VLQs couple directly with the SM gauge bosons, such mixings can affect significantly oblique parameters~\cite{Peskin:1990zt,Peskin:1991sw} and the $W$-boson mass. The focus of this work is to study the constraints of the oblique parameters on the VLQ extensions and their prediction of the $W$-boson mass.

This paper is organized as follows. In Secs. II and III, seven VLQ extensions and the calculation of the oblique parameters are introduced, respectively.
In Sec. IV, experimental and theoretical constraints on the extensions are scrutinized. It is emphasized that the global fit of the EWPOs plays an important role in this aspect. The capability of the VLQ extensions to explain the $W$-mass discrepancy is investigated in Sec. V, and conclusions are presented in Sec. VI.

\section{THE VLQ EXTENSIONS}
\label{sec:mix}

In this section, the key features of the VLQ extensions~\cite{Aguilar-Saavedra:2013qpa,Buchkremer:2013bha,Atre:2011ae,Chen:2017hak} are recapitulated. This brief introduction is restricted to the case where VLQ multiplets have renormalizable couplings to the SM Higgs field and, for the sake of simplicity, only one multiplet is involved in each extension. These multiplets are categorized into seven types by their charges of the SM gauge group $SU(3)_C \times SU(2)_L \times U(1)_Y$ in Table~\ref{tab:rp}, where the $SU(2)_L$ singlets $U$ and $D$, the doublets $Q_1$, $Q_5$, and $Q_7$, and the triplets $T_1$ and $T_2$ are distinguished by their hypercharge quantum number. Correspondingly, the component fields $T^\prime$, $B^\prime$, $X$, and $Y$ carry the electric charge of $2/3$, $-1/3$, $5/3$, and $-4/3$, respectively. Neglecting the small Yukawa couplings to the first two-generation quarks, the Higgs field ($H$) has the following interactions\footnote{In principle, the $Q_1$ multiplet may also couple with $H$ by the form $\bar{Q}_{1} H b_R$. To avoid unnecessary complexity, this interaction is neglected in this work, and this does not change the obtained conclusions much.}:
\begin{eqnarray}
-\mathcal{L}_{H} &=& Y_{t^\prime} \bar{q^\prime}_L \tilde{H} t_R^\prime + Y_{b^\prime} \bar{q^\prime}_L H b_R^\prime + \xi_{U} \bar{U} \tilde{H}^{\dagger} q_L^\prime + \xi_{D} \bar{D} \tilde{H}^{\dagger} q_L^\prime + \xi_{Q_1} \bar{Q}_{1} \tilde{H} t_R^\prime  + \xi_{Q_5} \bar{Q}_{5} \tilde{H} b_R^\prime  \nonumber \\
		& & + \xi_{Q_{7}} \bar{Q}_{7} H t_R^\prime +\frac{1}{2} \xi_{T_1} H^{\dagger} \tau \cdot \bar{T}_{1} q_L^\prime + \frac{1}{2} \xi_{T_2} \tilde{H}^{\dagger} \tau \cdot \bar{T}_{2} q_L^\prime + \text { H.c. },  \label{Yukawa}
\end{eqnarray}
where $q_L^\prime = (t^\prime,b^\prime)_L$ and $t^\prime_R/b^\prime_R$ are the left-handed and right-handed third-generation quark fields in the SM, respectively; $\tau$ denotes the Pauli matrix; and $\tilde{H} \equiv {\it{i}} \tau_2 H^\ast$. The coefficients $Y_i$ ($i=t^\prime,b^\prime$) and $\xi_j$ ($j=U$, $D$, $Q_1$, $Q_5$, $Q_7$, $T_1$, $T_2$) parametrize the Yukawa coupling strength, and only one of $\xi_j$ appears in this study of VLQ extensions.

%%%%%%%%%%%%%%%%%%%%%%%%%%%%%%%%%%%%%%%%%%%%%%%%

\begin{table}[tbp]
\centering
\begin{tabular}{c|ccccccc}
\hline
VLQ multiplet & $U$ & $D$ & $Q_1$ & $Q_5$ & $Q_7$ & $T_1$ & $T_2$  \\
\hline
Component fields & $T^\prime$ & $B^\prime$ & $(T^\prime,B^\prime)$ & $(B^\prime, Y)$ & $(X,T^\prime)$ & $(T^\prime,B^\prime,Y)$ & $(X,T^\prime,B^\prime)$ \\
$SU(3)_C$ &3 &3 &3 & 3 &3 &3 &3\\
$SU(2)_L$ &1 &1 &2 & 2 &2 &3 & 3 \\
$U(1)_Y$ & $2/3$  & $-1/3$ &$1/6$ & $-5/6$ &$7/6$ &$-1/3$&$2/3$ \\
\hline
\end{tabular}

\caption{Component fields of VLQ multiplets and their quantum number under the SM gauge groups. \label{tab:rp}}
\end{table}

The interactions in Eq.~(\ref{Yukawa}) imply that the VLQs mix with the third-generation quarks to form mass eigenstates. Assuming the presence of $T^\prime$ and $B^\prime$ fields,  the mass terms in the weak interaction bases $\Psi^t \equiv (t^\prime, T^\prime)$ and $\Psi^b \equiv (b^\prime, B^\prime)$ are given by
\begin{eqnarray}
&& -\mathcal{L}_{\rm mass} = \sum_{i,j=1,2} \left ( \bar{\Psi}_{Li}^t {\cal{M}}^t_{ij} \Psi^t_{Rj} + \bar{\Psi}_{Li}^b {\cal{M}}^b_{ij} \Psi^b_{Rj} \right ) + \text{H.c.} \label{ec:Lmass} \\
&=&  \left(\! \begin{array}{cc} \bar t^\prime_{L} & \bar T^\prime_{L} \end{array} \!\right)
	\left(\! \begin{array}{cc} Y_{t^\prime} \frac{v}{\sqrt 2} & {\cal{M}}^t_{12}  \\ {\cal{M}}^t_{21}  & M_0 \end{array} \!\right)
	\left(\! \begin{array}{c} t^\prime_{R}\\ T^\prime_{R} \end{array}
	\!\right) + \left(\! \begin{array}{cc} \bar b^\prime_{L} & \bar B^\prime_{L} \end{array} \!\right)
	\left(\! \begin{array}{cc} Y_{b^\prime} \frac{v}{\sqrt 2} & {\cal{M}}^b_{12}  \\ {\cal{M}}^b_{21}  & M_0 \end{array} \!\right)
	\left(\! \begin{array}{c} b^\prime_{R}\\ B^\prime_{R}  \end{array}
	\!\right) + \text{H.c.}. \nonumber
\end{eqnarray}
In this formula, the multiplet-universal bare mass $M_0$ is not related to the Higgs mechanism of the SM. Instead, it may be generated by a Yukawa coupling to a $SU(2)$-singlet scalar, such as the dilaton or radion field in Refs.~\cite{Abe:2012eu,Cao:2013cfa}, which acquires a vacuum expectation value much larger than $246~{\rm GeV}$. This mechanism can also solve the vacuum stability problem of the VLQ extensions and thus, it is of theoretical interest~\cite{Arsenault:2022xty}.

The mass matrices ${\cal{M}}^t$ and ${\cal{M}}^b$ in Eq.~(\ref{ec:Lmass}) can be diagonalized by biunitary transformations:
\begin{equation}
	V^{q \dagger}_L \, {\cal{M}}^q \, V^q_R = {\cal{M}}^q_\text{diag} = {\rm diag}(m_q, m_Q) \,,
\end{equation}
where $m_q$ and $m_Q$ are the masses of physical states with $(q,Q)=(t,T)$ or $(q,Q)=(b,B)$, respectively. Assuming no $CP$ violation in ${\cal{M}}^q$, the $2\times2$ unitary matrices $V_L^q$ and $V_R^q$ are parametrized by one mixing angle, respectively. That is
\begin{eqnarray}
V_{L,R}^q \equiv \left(\! \begin{array}{cc} \cos \theta^q_{L,R} & \sin \theta^q_{L,R}  \\  -\sin \theta^q_{L,R}  & \cos \theta^q_{L,R} \end{array} \!\right).
\end{eqnarray}
The following equations are obtained in the diagonalization:
\begin{eqnarray}
{\cal{M}}_{12}^{q} &=& -s_{R}^{q} c_{L}^{q} m_{q} + s_{L}^{q} c_{R}^{q} m_{Q}, \quad {\cal{M}}_{21}^{q} = -s_{L}^{q}c_{R}^{q}m_{q}+c_{L}^{q}s_{R}^{q} m_{Q}, \label{relation} \\
 M_0 &=& s_{L}^{q} s_{R}^{q} m_{q} + c_{L}^{q} c_{R}^{q} m_{Q}, \nonumber \\
({\cal{M}}_{21}^{q})^{2}+ M_0^2 &=& (s_{L}^{q})^{2}m_{q}^{2} + (c_{L}^{q})^{2}m_{Q}^{2}, \quad ({\cal{M}}_{12}^{q} )^2 + M_0^2 = (s_{R}^{q})^{2}m_{q}^{2} + (c_{R}^{q})^{2}m_{Q}^{2}, \nonumber \\
\frac{Y_{q^\prime} v}{\sqrt{2}}{\cal{M}}_{21}^{q}+ M_0 {\cal{M}}_{12}^{q} & = & s_{L}^{q} c_{L}^{q} \left ( m_{Q}^{2} - m_{q}^{2} \right ), \quad \frac{Y_{q^\prime} v}{\sqrt{2}}{\cal{M}}_{12}^{q} + M_0 {\cal{M}}_{21}^{q}  = s_{R}^{q} c_{R}^{q} \left ( m_{Q}^{2} - m_{q}^{2} \right ), \nonumber
\end{eqnarray}
where $s_{L,R}^q \equiv \sin \theta_{L,R}^q$ and $c_{L,R}^q \equiv \cos \theta_{L,R}^q$. These equations imply that $m_b$, $m_t$, and one of the four mixing angles can replace the Yukawa couplings of $Y_{b^\prime}$, $Y_{t^\prime}$, and $\xi_j$ as theoretical inputs, and $m_T$ or $m_B$ substitutes $M_0$ as an input. They decide the other mixing angles and masses.

Specifically, the VLQ extensions have the following distinct characteristics if the bottom quark mass is consistently neglected:
\begin{enumerate}
\item {\bf{$U$ extension}}: There is no $B^\prime$ field, and
\begin{eqnarray}
{\cal{M}}_{12}^t &=& \frac{\xi_{U} v}{\sqrt{2}},\quad {\cal{M}}_{21}^t = 0, \quad \tan \theta^t_R = \frac{m_t}{m_T} \tan \theta_L^t, \quad  M_0^2 = (s_L^t)^2 m_t^2 + (c_L^t)^2 m_T^2. \nonumber
\end{eqnarray}

\item {\bf{$D$ extension}}: There is no $T^\prime$ field, and
\begin{eqnarray}
{\cal{M}}_{12}^b &=& \frac{\xi_{D} v}{\sqrt{2}},\quad {\cal{M}}_{21}^b = 0, \quad \tan \theta^b_R = 0, \quad  M_0^2 = (c_L^b)^2 m_B^2. \nonumber
\end{eqnarray}

\item {\bf{$Q_1$ extension}}:
\begin{eqnarray}
{\cal{M}}_{12}^t &=& 0, \quad {\cal{M}}_{21}^t = \frac{\xi_{Q_1} v}{\sqrt{2}}, \quad {\cal{M}}_{12}^b = {\cal{M}}_{21}^b =  0,  \quad \theta_L^b = \theta_R^b =0, \nonumber \\
\tan \theta^t_L &=& \frac{m_t}{m_T} \tan \theta_R^t, \quad  m_B^2 = M_0^2 = (s_R^t)^2 m_t^2 + (c_R^t)^2 m_T^2.  \nonumber
\end{eqnarray}

\item {\bf{$Q_5$ extension}}: There is no $T^\prime$ field, and
\begin{eqnarray}
{\cal{M}}_{12}^b &=& 0, \quad {\cal{M}}_{21}^b = \frac{\xi_{Q_5} v}{\sqrt{2}}, \quad \tan \theta^b_L =0, \quad  m_Y^2 = M_0^2 = (c_R^b)^2 m_B^2.  \nonumber
\end{eqnarray}

\item {\bf{$Q_7$ extension}}: There is no $B^\prime$ field, and
\begin{eqnarray}
{\cal{M}}_{12}^t &=& 0, \quad {\cal{M}}_{21}^t = \frac{\xi_{Q_7} v}{\sqrt{2}}, \quad \tan \theta^t_L = \frac{m_t}{m_T} \tan \theta_R^t, \quad  m_X^2 = M_0^2 = (s_R^t)^2 m_t^2 + (c_R^t)^2 m_T^2.  \nonumber
\end{eqnarray}

\item {\bf{$T_1$ extension}}:
\begin{eqnarray}
{\cal{M}}_{12}^t &=& \xi_{T_1} v, \quad {\cal{M}}_{21}^t = 0, \quad {\cal{M}}_{12}^b = \frac{\xi_{T_1} v}{\sqrt{2}}, \quad {\cal{M}}_{21}^b = 0, \nonumber \\
\tan \theta^t_R &=& \frac{m_t}{m_T} \tan \theta_L^t, \quad \theta_R^b = 0, \quad  m_Y^2 = M_0^2 = (s_L^t)^2 m_t^2 + (c_L^t)^2 m_T^2,    \nonumber  \\
m_B^2 &=& \frac{1}{2} \left [ (c_L^t)^2 m_T^2 + (c_R^t)^2 m_T^2 +  (s_L^t)^2 m_t^2 + (s_R^t)^2 m_t^2 \right ], \quad c_L^b = \frac{m_Y}{m_B}. \nonumber
\end{eqnarray}

\item {\bf{$T_2$ extension}}:
\begin{eqnarray}
{\cal{M}}_{12}^t &=& \frac{\xi_{T_2} v}{\sqrt{2}}, \quad {\cal{M}}_{21}^t = 0, \quad {\cal{M}}_{12}^b = \xi_{T_2} v, \quad {\cal{M}}_{21}^b = 0, \nonumber \\
\tan \theta^t_R &=& \frac{m_t}{m_T} \tan \theta_L^t, \quad \theta_R^b = 0, \quad  m_X^2 = M_0^2 = (s_L^t)^2 m_t^2 + (c_L^t)^2 m_T^2,    \nonumber  \\
m_B^2 &=& 2 (c_R^t)^2 m_T^2 - (c_L^t)^2 m_T^2 +  2 (s_R^t)^2 m_t^2 - (s_L^t)^2 m_t^2, \quad c_L^b = \frac{m_X}{m_B}. \nonumber
\end{eqnarray}

\end{enumerate}
Note that the Yukawa interaction in Eq.~(\ref{Yukawa}) for the $T_1$ and $T_2$ extensions determines both the $t^\prime-T^\prime$ mixing and the $b^\prime-B^\prime$ mixing simultaneously. As a result, the mixing angles in the top and bottom sectors are related by $s_L^t \simeq \sqrt{2} s_L^b$ for the $T_1$ extension and $s_L^t \simeq  s_L^b/\sqrt{2}$ for the $T_2$ extension. Given that a sizable $s_L^b$ can significantly alter the $Z b \bar{b}$ couplings of the SM, the mixings are tightly limited (see the discussion later).

In the following, $s_L^t$ and $m_T$ for the $U$, $T_1$, and $T_2$ extensions, $s_L^b$ and $m_B$ for the $D$ extension, $s_R^t$ and $m_T$ for the $Q_1$ and $Q_7$ extensions, and $s_R^b$ and $m_B$ for the $Q_5$ extension are used as theoretical inputs to study the impacts of the VLQ extensions on the oblique parameters and $W$-boson mass.

\section{OBLIQUE PARAMETERS}

In the VLQ extensions, the physical states $q$ and $Q$ contribute to the transverse component of the vacuum polarization for the gauge bosons in the SM through loop Feynman diagrams. This effect is formulated as follows~\cite{Cao:2008rc}
\begin{eqnarray}
\Sigma_{V^{\prime} V} \left(p^{2}\right) & = & \sum_{i,j=q,Q} \frac{2}{16 \pi^{2}}\left\{( g _ { L } ^ { \overline { \psi } _ { j } \psi _ { i } V ^ { \prime } } g _ { L } ^ { \overline { \psi } _ { i } \psi _ { j } V ^ { * } } + g _ { R } ^ { \overline { \psi } _ { j } \psi _ { i } V ^ { \prime } } g _ { R } ^ { \overline { \psi } _ { i } \psi _ { j } V ^ { * } } ) ( 2 p ^ { 2 } B _ { 3 } - B _ { 4 } ) \left(p, m_{\psi_{i}}, m_{\psi_{j}})\right.\right. \nonumber \\
& & \left.+\left(g_{L}^{\bar{\psi}_{j} \psi_{i} V^{\prime}} g_{R}^{\bar{\psi}_{i} \psi_{j} V^{*}}+g_{R}^{\bar{\psi}_{j} \psi_{i} V^{\prime}} g_{L}^{\bar{\psi}_{i} \psi_{j} V^{*}}\right) m_{\psi_{i}} m_{\psi_{j}} B_{0}\left(p, m_{\psi_{i}}, m_{\psi_{j}}\right)\right\}, \label{Self-energy}
\end{eqnarray}
where $p^2$ is the squared momentum of the incoming gauge boson, $\psi_i$ and $\psi_j$ denote the quark states entering the loop, $g_L^{\bar{\psi}_j \psi_i V}$ and $g_R^{\bar{\psi}_j \psi_i V}$ are their chiral couplings to the vector boson $V$ presented in the Appendix of this work, and contributions from different configurations of the states are summed.\footnote{Note that in this work, $\Sigma_{V^{\prime} V}$  is defined as the coefficient of the $- g_{\mu \nu}$ term in the vacuum-polarization tensor of the gauge bosons ~\cite{Bohm:1986rj,Hollik:1988ii}. It differs from the quantity $A_{V V^\prime}$ defined in Refs.~\cite{Peskin:1990zt,Peskin:1991sw} by a minus sign.} The loop functions $B_3$ and $B_4$ are related to the standard loop functions $B_1$ and $B_{21}$ by~\cite{Cao:2008rc}
\begin{eqnarray}
B_{3}\left(p, m_{1}, m_{2}\right)&=&-B_{1}\left(p, m_{1}, m_{2}\right)-B_{21}\left(p, m_{1}, m_{2}\right), \nonumber \\
B_{4}\left(p, m_{1}, m_{2}\right)&=&-m_{1}^{2} B_{1}\left(p, m_{2}, m_{1}\right)-m_{2}^{2} B_{1}\left(p, m_{1}, m_{2}\right).
\end{eqnarray}
In addition, the singlet scalar field responsible for the bare mass $M_0$ in Eq.~(\ref{ec:Lmass}) can contribute to the vacuum polarization. Such a contribution, however, is induced by the mixing of the scalar and the SM Higgs and usually neglected. This is because the LHC Higgs data as well as the vacuum stability constrain the mixing to make it small (see the studies in Refs.~\cite{Abe:2012eu,Cao:2013cfa} and Ref.~\cite{Arsenault:2022xty}, respectively), and also because the scalar-mediated loops are usually less important than fermionic loops in contributing to the vacuum polarization.

At the point $p^2=0$, $\Sigma_{V^{\prime} V}$ in Eq.(\ref{Self-energy}) can be simplified as follows
\begin{eqnarray}
\Sigma_{V^{\prime} V} \left( 0 \right) & = & \sum_{i,j} \frac{2}{16 \pi^{2}}\left\{( g _ { L } ^ { \overline { \psi } _ { j } \psi _ { i } V ^ { \prime } } g _ { L } ^ { \overline { \psi } _ { i } \psi _ { j } V ^ { * } } + g _ { R } ^ { \overline { \psi } _ { j } \psi _ { i } V ^ { \prime } } g _ { R } ^ { \overline { \psi } _ { i } \psi _ { j } V ^ { * } } ) F_1 \left ( m_{\psi_{i}}, m_{\psi_{j}} \right) \right. \nonumber \\
& & \left.+\left(g_{L}^{\bar{\psi}_{j} \psi_{i} V^{\prime}} g_{R}^{\bar{\psi}_{i} \psi_{j} V^{*}}+g_{R}^{\bar{\psi}_{j} \psi_{i} V^{\prime}} g_{L}^{\bar{\psi}_{i} \psi_{j} V^{*}}\right) m_{\psi_{i}} m_{\psi_{j}} F_2 \left(m_{\psi_{i}}, m_{\psi_{j}}\right)\right\}, \\
\frac{\partial \Sigma_{V^\prime V} \left( p^2 \right)}
{\partial p^2}|_{p^2=0} & = & \sum_{i,j} \frac{2}{16 \pi^{2}}\left\{( g _ { L } ^ { \overline { \psi } _ { j } \psi _ { i } V ^ { \prime } } g _ { L } ^ { \overline { \psi } _ { i } \psi _ { j } V ^ { * } } + g _ { R } ^ { \overline { \psi } _ { j } \psi _ { i } V ^ { \prime } } g _ { R } ^ { \overline { \psi } _ { i } \psi _ { j } V ^ { * } } ) F_3 \left (m_{\psi_{i}}, m_{\psi_{j}}) \right.\right. \nonumber \\
& & \left.+\left(g_{L}^{\bar{\psi}_{j} \psi_{i} V^{\prime}} g_{R}^{\bar{\psi}_{i} \psi_{j} V^{*}}+g_{R}^{\bar{\psi}_{j} \psi_{i} V^{\prime}} g_{L}^{\bar{\psi}_{i} \psi_{j} V^{*}}\right) m_{\psi_{i}} m_{\psi_{j}} F_4 \left(m_{\psi_{i}}, m_{\psi_{j}}\right)\right\},
\end{eqnarray}
where
\begin{eqnarray}
F_{1}\left(m_{1}, m_{2}\right) &=& \frac{m_{1}^{2}}{2} \ln \frac{m_{2}^{2}}{\mu^{2}}+ \frac{m_{2}^{2}}{2} \ln \frac{m_{1}^{2}}{\mu^{2}}-\frac{m_{1}^{2}+m_{2}^{2}}{4} - \frac{m_{1}^{2}-m_{2}^{2}}{2} \ln \frac{m_{2}^{2}}{m_{1}^{2}}-\frac{m_{1}^{2} m_{2}^{2}}{2\left(m_{1}^{2}-m_{2}^{2}\right)} \ln \frac{m_{2}^{2}}{m_{1}^{2}}, \nonumber \\
F_{2}\left(m_{1}, m_{2}\right) &=& -\frac{1}{2}\left(\ln \frac{m_{1}^{2}}{\mu^{2}}+\ln \frac{m_{2}^{2}}{\mu^{2}}\right)+1+\frac{m_{1}^{2}+m_{2}^{2}}{2\left(m_{1}^{2}-m_{2}^{2}\right)} \ln \frac{m_{2}^{2}}{m_{1}^{2}}, \nonumber \\
F_{3}\left(m_{1}, m_{2}\right) &=& -\frac{1}{6}\left ( \ln \frac{m_{1}^{2}}{\mu^{2}}+\ln \frac{m_{2}^{2}}{\mu^{2}}\right ) +\frac{1}{9}-\frac{2 m_{1}^{2} m_{2}^{2}}{3\left(m_{1}^{2}-m_{2}^{2}\right)^{2}} \nonumber \\
& & +\frac{m_{1}^{6}+m_{2}^{6}-3 m_{1}^{2} m_{2}^{2}\left(m_{1}^{2}+m_{2}^{2}\right)}{6\left(m_{1}^{2}-m_{2}^{2}\right)^{3}} \ln \frac{m_{2}^{2}}{m_{1}^{2}}, \nonumber \\
F_{4}\left(m_{1}, m_{2}\right) &= & \frac{m_{1}^{4}}{2\left(m_{1}^{2}-m_{2}^{2}\right)^{3}}-\frac{m_{2}^{4}}{2\left(m_{1}^{2}-m_{2}^{2}\right)^{3}}+\frac{m_{1}^{2} m_{2}^{2}}{\left(m_{1}^{2}-m_{2}^{2}\right)^{3}} \ln \frac{m_{2}^{2}}{ m_{1}^{2}},  \label{fun}
\end{eqnarray}
with $\mu$ denoting the renormalization scale. Note that $F_i (m_1,m_2)$ is symmetric under the exchange of $m_1$ and $m_2$, and if $m_2 \simeq m_1$, then they can be approximated by
\begin{eqnarray}
F_1(m_1,m_2) &\simeq& \frac{m_1^2}{2} \ln \frac{m_2^2}{\mu^2} + \frac{m_2^2}{2} \ln \frac{m_1^2}{\mu^2} + \frac{5}{12} \frac{(m_2^2- m_1^2)^2}{m_1^2}, \nonumber \\
F_2(m_1,m_2) &\simeq& -\frac{1}{2} \left ( \ln \frac{m_1^2}{\mu^2} + \ln \frac{m_2^2}{\mu^2} \right ) - \frac{1}{12} \frac{(m_2^2 - m_1^2)^2}{m_1^4}, \nonumber\\
F_3(m_1,m_2) &\simeq& -\frac{1}{6} \left ( \ln \frac{m_1^2}{\mu^2} + \ln \frac{m_2^2}{\mu^2} \right ) - \frac{1}{6}, \quad F_4(m_1,m_2) \simeq \frac{1}{4 m_1^2} - \frac{m_2^2}{12 m_1^4}, \nonumber
\end{eqnarray}
when $\mu$ is set at the electroweak scale.

The oblique parameters $S$, $T$, and $U$ are defined in terms of the weak isospin current $J^\mu_{1,2,3}$ and the electromagnetic current $J^\mu_Q = J^\mu_3 + J^\mu_Y$  by their vacuum-polarization amplitudes~\cite{Peskin:1990zt,Peskin:1991sw}
\begin{eqnarray}
S &\equiv& - \frac{16 \pi}{m_Z^2} \left \{ \Sigma_{33} (m_Z^2) - \Sigma_{33} (0) - \Sigma_{3Q} (m_Z^2) \right \} \nonumber \\
&=&  \frac{16 \pi}{m_Z^2} \left \{ \Sigma_{3Y} (m_Z^2) - \Sigma_{3Y} (0) \right \},  \label{S} \\
T &\equiv& \frac{4 \pi}{s_W^2 c_W^2 m_Z^2} \left \{ \Sigma_{33} (0) - \Sigma_{11} (0) \right \},  \label{T} \\
U &\equiv& \frac{16 \pi}{m_Z^2} \left \{ \Sigma_{33} (m_Z^2) - \Sigma_{33} (0) \right \} -  \frac{16 \pi}{m_W^2} \left \{ \Sigma_{11} (m_Z^2) - \Sigma_{11} (0) \right \},  \label{U}
\end{eqnarray}
where $s_W \equiv \sin \theta_W$ and $c_W \equiv \cos \theta_W$ are the sine and cosine, respectively, of the weak mixing angle $\theta_W$, and $m_W$ and $m_Z$ denote the mass for $W$ and $Z$ boson, respectively. Evidently, $T$ and $U$ receive nonzero contributions from the violation of the weak isospin, and they are finite because of the weak
isospin symmetric nature of the divergence terms. $S$ originates from the mixing between the weak hypercharge and the
third component of the weak isospin, which results from the spontaneous symmetry breakdown. It involves only soft operators and thus possesses no divergences~\cite{Peskin:1990zt,Peskin:1991sw,Grinstein:1991cd}. Note that the above definitions are complete in the sense of including contributions from
the SM and any possible new physics. As the vacuum-polarization amplitudes $\Sigma_{ij}$ ($i,j=1,2,3,Q,Y$) receive contributions from different sources additively at the one-loop level, alternative $S$, $T$, and $U$ can be defined by Eqs.~(\ref{S}-\ref{U}), respectively, with
$\Sigma^{\rm new} (p^2) \equiv \Sigma^{\rm NP}(p^2) - \Sigma^{\rm SM} (p^2)$. In this case, $S$, $T$, and $U$ only contain new physics effects.

Given that the $Z$-boson current is equal to $\frac{e}{s_W c_W} ( J^\mu_3 - s_W^2 J^\mu_Q )$, with $e$ related to the fine-structure constant $\alpha$ by $e^2 \equiv 4 \pi \alpha$, the oblique parameters can be reexpressed by the vacuum polarizations of the SM gauge bosons as
\begin{eqnarray}
\frac{\alpha}{4 s_W^2 c_W^2}\, S &=&
- \frac{\Sigma_{ZZ}^{\rm new} \left( m_Z^2 \right) - \Sigma_{ZZ}^{\rm new} \left( 0 \right)}{m_Z^2}
+ \left. \frac{\partial \Sigma_{\gamma \gamma}^{\rm new} \left( p^2 \right)}
{\partial p^2} \right|_{p^2=0}
+ \frac{c_W^2 - s_W^2}{c_W s_W}
\left. \frac{\partial \Sigma_{\gamma Z}^{\rm new} \left( p^2 \right)}
{\partial p^2} \right|_{p^2=0} \nonumber \\
&\simeq& \left. - \frac{\partial \Sigma_{Z Z}^{\rm new} \left( p^2 \right)}
{\partial p^2} \right|_{p^2=0} + \left. \frac{\partial \Sigma_{\gamma \gamma}^{\rm new} \left( p^2 \right)}
{\partial p^2} \right|_{p^2=0}
+ \frac{c_W^2 - s_W^2}{c_W s_W}
\left. \frac{\partial \Sigma_{\gamma Z}^{\rm new} \left( p^2 \right)}
{\partial p^2} \right|_{p^2=0}, \\
\alpha T &=&
\frac{\Sigma_{ZZ}^{\rm new} \left( 0 \right)}{m_Z^2}
-\frac{\Sigma_{WW}^{\rm new} \left( 0 \right)}{m_W^2}, \\
\frac{\alpha}{4 s_W^2}\, U &=& - \frac{\Sigma_{WW}^{\rm new} \left( m_W^2 \right) - \Sigma_{WW}^{\rm new} \left( 0 \right)}{m_W^2}
+ c_W^2\,
\frac{\Sigma_{ZZ}^{\rm new} \left( m_Z^2 \right) - \Sigma_{ZZ}^{\rm new} \left( 0 \right)}{m_Z^2}  \nonumber \\
& & + s_W^2
\left. \frac{\partial \Sigma_{\gamma \gamma}^{\rm new} \left( p^2 \right)}
{\partial p^2} \right|_{p^2=0}
+ 2 c_W s_W
\left. \frac{\partial \Sigma_{\gamma Z}^{\rm new} \left( p^2 \right)}
{\partial p^2} \right|_{p^2=0}  \nonumber \\
&\simeq&  - \frac{\partial \Sigma_{WW}^{\rm new} \left( p^2 \right)}
{\partial p^2}|_{p^2=0} + c_W^2 \frac{\partial \Sigma_{ZZ}^{\rm new} \left( p^2 \right)}
{\partial p^2}|_{p^2=0} + s_W^2 \frac{\partial \Sigma_{\gamma \gamma}^{\rm new} \left( p^2 \right)}
{\partial p^2}|_{p^2=0} \nonumber \\
& & + 2 c_W s_W \frac{\partial \Sigma_{\gamma Z}^{\rm new} \left( p^2 \right)}
{\partial p^2}|_{p^2=0}.
\end{eqnarray}
In this study, the latter formulas are used in the calculation of the oblique parameters. We check that the specific forms of the quark mass matrix listed at the end of the last section play a vital role in canceling out the UV divergence of the loop functions. As a result, the oblique parameters are free of the divergence. With the aid of the oblique  parameters, the new physics correction to $W$-boson mass is given by~\cite{Bohm:1986rj,Hollik:1988ii,Maksymyk:1993zm,Burgess:1993mg}
\begin{eqnarray}
\frac{\delta m_W}{m_W} = \frac{\alpha}{2 (c_W^2 - s_W^2)} \left (-\frac{1}{2} S + c_W^2 T + \frac{c_W^2-s_W^2}{4 s_W^2} U \right ).  \label{m_W}
\end{eqnarray}

\section{EXPERIMENTAL AND THEORETICAL CONSTRAINTS}

So far, the VLQ extensions have been restricted nontrivially by collider data and theoretical preferences. In attempting to interpret the mass anomaly, the following constraints should be considered.

\subsection{Constraints from the EWPOs}

The theories under study affect the EWPOs not only via the vacuum polarization of the gauge bosons, which is parametrized by the oblique parameters, but also by modifying the $Z b \bar{b}$ couplings at the tree level by the $b^\prime-B^\prime$ mixing. A systematic study of the EWPOs' constraints includes formulating the EWPOs as linear functions of $S$, $T$, $U$, $\delta g_L^b$, and $\delta g_R^b$~\cite{Ciuchini:2013pca}, where $\delta g_L^b$ and $\delta g_R^b$ denote the correction to the $Z b_L \bar{b}_L$ and $Z b_R \bar{b}_R$ couplings, respectively, and sequentially fitting them to low-energy experimental measurements~\cite{deBlas:2016ojx}. It is found that the correlation between the oblique parameters and $\delta g_{L,R}^b$ is much weaker than those among the oblique parameters and between $\delta g_L^b$ and $\delta g_R^b$~\cite{deBlas:2016ojx}. This conclusion implies that as a good approximation, one may fit separately the oblique parameters to the EWPOs and $\delta g_L^b$ and $\delta g_R^b$ to the measurements $R_b^{\rm exp}$, $A_{\rm FB}^{b, \rm exp}$, $A_b^{\rm exp}$ and $R_c^{\rm exp}$  in implementing the constraints. This strategy was adopted in Ref.~\cite{Aguilar-Saavedra:2013qpa}.

\begin{table}[t]
	\begin{center}
		\begin{tabular}{c|c|c|c|c|c|c|c}
			\hline
			& $U$ & $D$ &$Q_1$ & $Q_5$ & $Q_7$ & $T_1$ & $T_2$ \\
			\hline
			${\rm BR}(T \to b W)$ & 50\%  & $\cdots$ & $< 1\%$ & $\cdots$ & $< 1\%$ & $< 1\%$ & 50\% \\
			\hline
			${\rm BR}(T \to t Z)$ & 25\% & $\cdots$ & 50\% & $\cdots$ & 50\% & 50\% & 25\% \\
			\hline
			${\rm BR}(T \to t H)$ &25\% & $\cdots$ & 50\% & $\cdots$ & 50\% & 50\% & 25\% \\
			\hline
			${\rm BR}(B \to t W)$ & $\cdots$ & 50\% & 100\% & $< 1\%$  & $\cdots$ & 50\% & $< 1\%$ \\
			\hline
			${\rm BR}(B \to b Z)$ & $\cdots$ & 25\% & $< 1\%$ & 50\% & $\cdots$ & 25\% & 50\% \\
			\hline
			${\rm BR}(B \to b H)$ & $\cdots$ & 25\% & $< 1\%$ & 50\% & $\cdots$ & 25\% & 50\% \\
			\hline
		\end{tabular}
		\caption{Approximate decay branching ratios of the states $T$ and $B$ in different VLQ extensions with the Lagrangian given in Eq.~(\ref{Yukawa}).  \label{bran}}
	\end{center}
\end{table}

In this study, results from the latest global fit of the oblique parameters in Ref.~\cite{deBlas:2022hdk} and the $Z b \bar{b}$ coupling fit in Ref.~\cite{Aguilar-Saavedra:2013qpa} are used to constrain the VLQ theories. The former fit includes the recent CMS measurement of top quark mass and the latest CDF measurement of $W$-boson mass to obtain the preference of the oblique parameters. It considers the standard average scenario and the conservative average scenario. The first scenario adopts the averaged (or combined) value of $m_t$ and $m_W$ from different experimental collaborations, $m_t^{\rm Cb} =171.79 \pm 0.38~{\rm GeV}$ and $m_W^{\rm Cb} = 80.4133 \pm 0.008~{\rm GeV}$, as the fit inputs.  The second scenario, however, considers the tensions among individual measurements of $m_t$ and those of $m_W$ (e.g., the tensions between the measurement of $m_W$ in Ref.~\cite{CDF:2022hxs} and Refs.~\cite{ATLAS:2017rzl,LHCb:2021bjt}), and thus adopts larger error bars, that is, $m_t^{\rm Cb} =171.79 \pm 1.0~{\rm GeV}$ and $m_W^{\rm Cb} = 80.4133 \pm 0.015~{\rm GeV}$, in performing the fit. Correspondingly, the standard average scenario imposes much stronger constraints than the conservative average scenario~\cite{deBlas:2022hdk}. In numerical calculation, the $\chi^2$ function for the oblique parameter fit is required to be less than 3.53 and 8.02 for three degrees of freedom to compute the $1\sigma$ and $2\sigma$ confidence level limits, respectively. Concerning the $Z b \bar{b}$ coupling fit, note that there has been no significant improvement on the measurements of $R_b$, $A_{\rm FB}^{b}$, $A_b$ and $R_c$ in recent years. Thus, the $95\%$ confidence level bounds in Ref.~\cite{Aguilar-Saavedra:2013qpa} can be directly used in this study. After comparing the constraints from the two separate fits, it is concluded that the oblique fit limits the parameter space of the $U$, $Q_1$, $Q_5$, and $Q_7$ extensions more tightly than the $Z b\bar{b}$ coupling fit, and the situation is reversed for the $D$, $T_1$, and $T_2$ extensions. The underlying reason is that the $Z b \bar{b}$ couplings are different from their SM prediction only at the loop level for the $U$, $Q_1$, and $Q_7$ extensions, while they are modified at the tree level for the $D$, $T_1$, and $T_2$ extensions. The situation of the $Q_5$ extension is somewhat subtle, in that it provides a positive correction to $\delta g_R^b$ at the tree level, and such a $\delta g_R^b$ is favored by experimental data. Consequently, the constraints from the $Z b \bar{b}$ coupling fit on the $Q_5$ extension is relaxed greatly~\cite{Aguilar-Saavedra:2013qpa}.

\subsection{The LHC search for VLQs}

In the VLQ extensions, the VLQ-dominated state $Q$ decays into the $t^\prime$- and $b^\prime$-dominated states $t$ and $b$ by the following channels: $T \to b W^+, t Z, t H$, $B \to t W^-, b Z, b H$, $X \to t W^+$, and $Y \to b W^-$. Given the Lagrangian in Eq.~(\ref{Yukawa}), the decay branching ratios are roughly fixed when $m_Q \gg v$~\cite{Aguilar-Saavedra:2013qpa}, and they are presented in Table~\ref{bran}. This feature is utilized to search for $Q$ at the LHC with $\sqrt{s} = 13~{\rm TeV}$.

\subsubsection{Single VLQ productions}

The state $T$ may be singly produced at the LHC by the parton processes $q g \to q^\prime (W^+ b) \bar{b} \to q^\prime (T) \bar{b}$, and its cross section is proportional to $\sin^2 \theta_L^t$ for the $U$ extension. The following analyses are conducted for this production:
\begin{itemize}
\item Search for the decays $T \to t H \to (b W^+) (b \bar{b}) \to (b l^+ \nu) (b \bar{b})$ and $ T \to t Z \to  (b l^+ \nu) (q \bar{q})$~ \cite{ATLAS:2021ddx,CMS:2021sqj}. Assuming $\sin \theta_L^t \geq 0.5$ $(0.41)$, a lower bound of $m_T \geq 1.8 {\rm TeV}$ $(1.6 {\rm TeV})$ is obtained, with an integrated luminosity of $139~{\rm fb^{-1}}$.
\item Search for the decay $T \to t W^+ \to (b l^+ \nu) (q \bar{q}^\prime)$~\cite{ATLAS:2018dyh}. It is found that $m_T \geq 800~{\rm GeV}$ for $\sin \theta_L^t = 0.18$, with an integrated luminosity of $36.1~{\rm fb^{-1}}$.
\end{itemize}

Note that these bounds cannot be applied to this study because, as indicated below, $\sin \theta_L^t$ is tightly limited.

\subsubsection{VLQ pair productions}

The state $Q$ may also be pair produced at the LHC via the QCD process $g g \to Q \bar{Q}$. The cross section of this process is independent of any mixing angles even though it is suppressed by the phase space when $Q$ is massive. This feature, together with the branching ratios in Table~\ref{bran}, enables the LHC search results to be directly applied to this study.

Specifically, the following conclusions are drawn based on analyses of the pair productions:
\begin{itemize}
\item The lower bounds of $m_T \geq 1.31~{\rm TeV}$ for the $U$ extension, $m_B \geq 1.22~{\rm TeV}$ for the $D$ extension, and $m_T, m_B \geq 1.37~{\rm TeV}$ for the $Q_1$ extension are obtained after combining all the decay channels of $T$ and $B$ with $36.1~{\rm fb^{-1}}$ integrated luminosity~\cite{ATLAS:2018ziw}.
\item The lower bounds of $m_T \geq 1.27~{\rm TeV}$ for the $U$ extension, $m_B \geq 1.20~{\rm TeV}$ for the $D$ extension, $m_T \geq 1.46~{\rm TeV}$ and $m_B \geq 1.32~{\rm TeV}$ for the $Q_1$ extension, and $m_T \geq 1.46~{\rm TeV}$ for the $Q_7$ extension are obtained with $139~{\rm fb^{-1}}$ integrated luminosity~\cite{ATLAS:2021ibc}. The analyses require one of $T$ ($B$) to decay by $T \to t Z$ ($B \to b Z$) with $Z \to \ell \bar{\ell}$ and concentrate on two topologies: exactly the two-lepton final state and at least, the three-lepton final states.
\item The lower bounds of $m_X \geq 1.33~{\rm TeV}$ and  $m_X \geq 1.30~{\rm TeV}$ are set for the cases of a purely right-handed and left-handed coupling to $W$ boson, respectively, with $35.9~{\rm fb^{-1}}$ integrated luminosity~\cite{CMS:2018ubm}. The analyses require, at least, one of the four $W$ bosons from $X \bar{X} \to t W^+ \bar{t} W^- \to (2 b) (4 W)$ to decay leptonically and focus on the single-lepton final state and same-sign dilepton final state.
\end{itemize}
Given the experimental status, a conservative bound of $m_T, m_B \geq 1.27~{\rm TeV}$ is used to present the results of this work. It is emphasized that a higher bound does not change the conclusions much.

\subsection{Measurements of the top quark property}

As shown in the Appendix, the couplings of the top quark to the SM gauge bosons are modified at the tree level in the VLQ extensions, which affect top quark properties such as the width and the single top production rate at the LHC~\cite{Aguilar-Saavedra:2013qpa}. In the $U$ extension, the modification is proportional to $\sin^2 \theta_L^t$ in leading approximation and thus at the order of $1\%$ for $\sin \theta_L^t \sim 0.1$, which is preferred to explain the mass anomaly (see the following discussion). The modification is much below the precision of current experimental measurements and has no impact on this study. Similar conclusions are obtained for the other extensions.

\subsection{The Higgs data collected at the LHC}

The Lagrangian in Eq.(\ref{Yukawa}) changes the Higgs couplings to the third-generation quarks. It also introduces $H \bar{Q} Q$ interaction. In the $U$ extension, both the modification of the $H \bar{t} t$ coupling from its SM prediction and the new Yukawa coupling coefficient $Y_{T}$ are proportional to $\sin^2 \theta_L^t$~\cite{Aguilar-Saavedra:2013qpa}. Consequently, the modifications of the $H \gamma \gamma$ and $H g g$ couplings are proportional to $\sin^2 \theta_L^t$, too~\cite{Djouadi:2005gj}. For $\sin \theta_L^t \sim 0.1$, the deviations of these couplings from their SM values are at the order of $1\%$ even when $m_T$ and $m_t$ are comparable. However, they only slightly alter the result of the SM Higgs fit to the LHC data. This conclusion holds for other extensions.

\subsection{Perturbativity}

In the VLQ extensions, the perturbativity of a theory at the electroweak scale requires $\xi_j$ in Eq.(\ref{Yukawa}) to satisfy $|\xi_j| \lesssim \sqrt{4 \pi}$. This condition sets an upper bound on the mixing angles. For example, $\tan 2 \theta_L^t$ in the $U$ extension is given by~\cite{Aguilar-Saavedra:2013qpa}
\begin{eqnarray}
\tan 2 \theta_L^t & = & \frac{\sqrt{2} |\xi_U| v M_0}{M_0^2 - y_{t^\prime}^2 v^2/2 - \xi_U^2 v^2/2}.  \label{slt}
\end{eqnarray}
In the limit $M_0 \gg v$, $m_T \simeq M_0$, and the perturbativity implies that
\begin{eqnarray}
\sin \theta_L^t \lesssim \sqrt{2 \pi} v/m_T, \quad {\rm or\ equivalently,}  \quad \sin \theta_L^t \lesssim 0.62 \times \left ( \frac{\rm TeV}{m_T} \right ).
\end{eqnarray}
Similar requirements can be set for $\sin \theta_L^b$ in the $D$ extension, $\sin \theta_R^t$ in the $Q_1$ and $Q_7$ extensions, $\sin \theta_R^b$ in the $Q_5$ extension, and $\sin \theta_L^t$ in the $T_1$ and $T_2$ extensions. As shown below, most of them are significantly weaker than the oblique parameters in limiting the parameter space of the VLQ theories for $m_T \lesssim 2~{\rm TeV}$.

In addition to the perturbativity, a moderately large $\xi_j$ can decrease significantly the Higgs quartic coupling at high energies to exacerbate the vacuum stability problem of the SM. Avoiding such a problem imposes special requirements on the theory~\cite{Arsenault:2022xty,Gopalakrishna:2018uxn}. This issue will be discussed at the end of Sec. V.

%\vspace{-0.5cm}

\section{$W$-BOSON MASS IN THE VLQ MODELS}

In this section, the predictions of $W$-boson mass in the VLQ extensions are presented. The oblique parameters are calculated to limit the parameter space of the theories. In addition, other important supplementary constraints are considered.

%\vspace{-0.3cm}

\subsection{$U$ extension}

\begin{figure}[t]	
	\centering
	\includegraphics[scale=0.33]{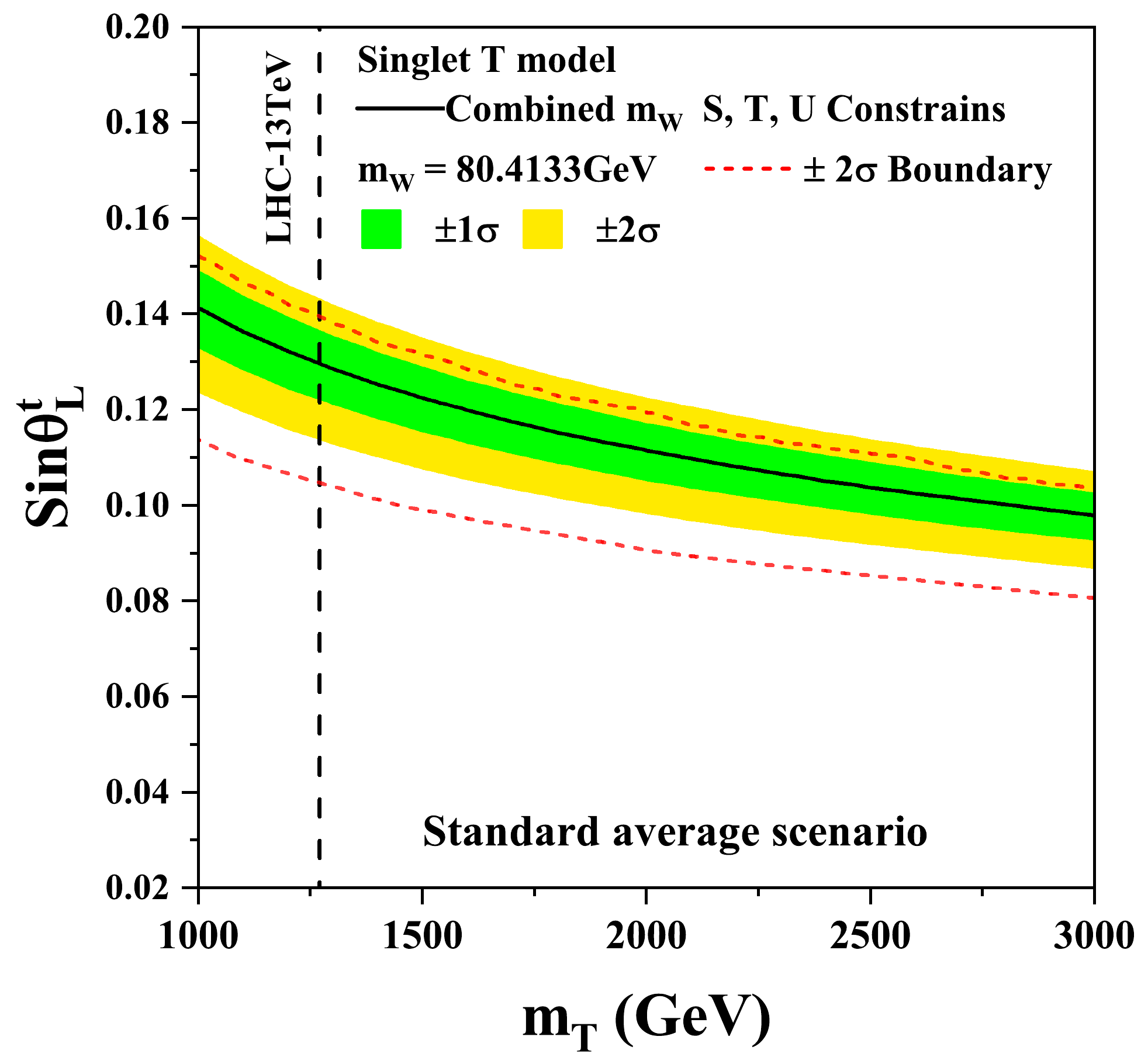}
	\includegraphics[scale=0.33]{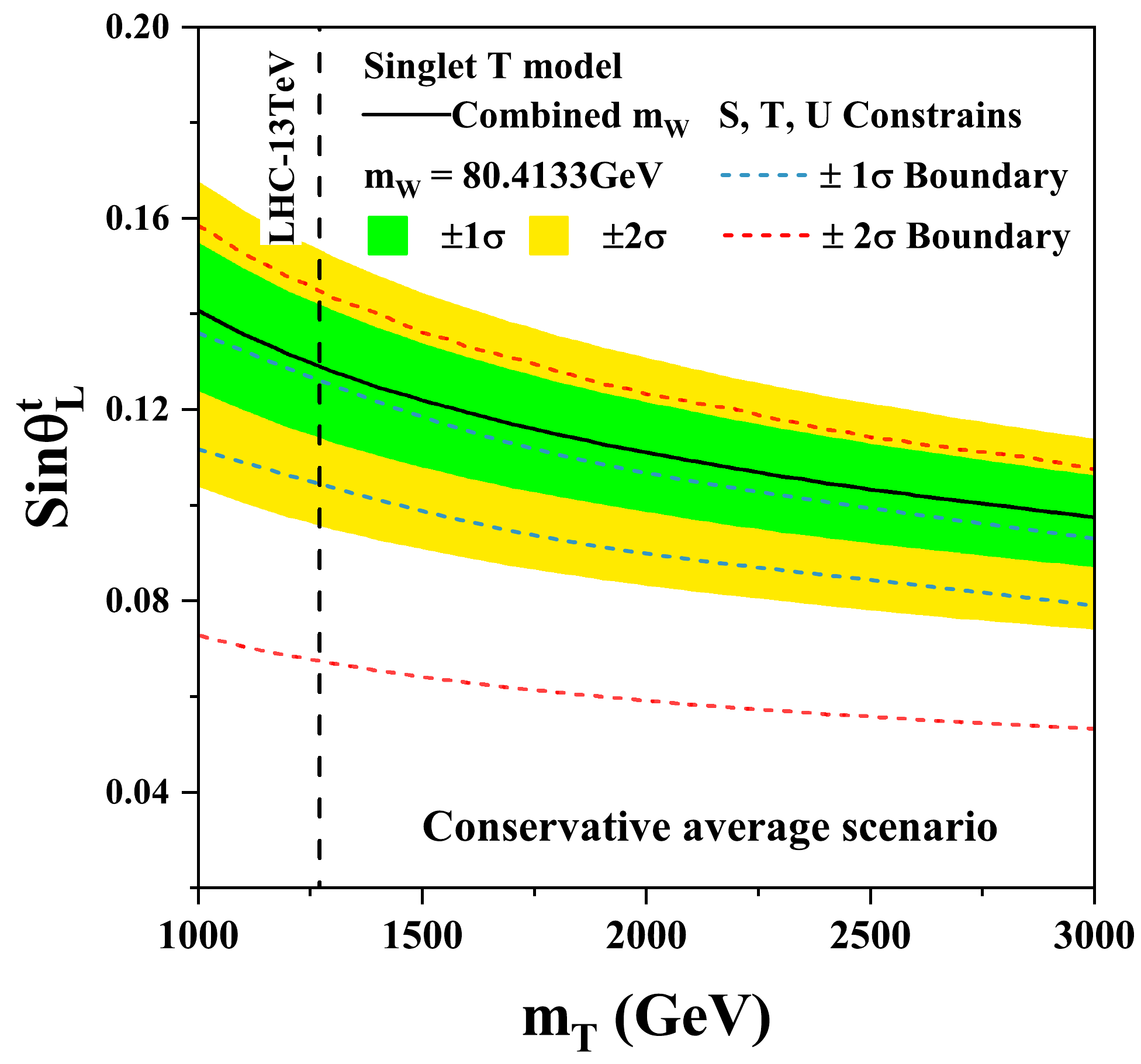}

\vspace{-0.4cm}

	\caption{Parameter space predicting the combined experimental value of $m_W$ in the $U$ extension, which is projected onto the $m_T-\sin \theta_L^t$ plane. The black contour corresponds to  $ m_W = 80.4133~{\rm GeV}$, the central value of $m_W^{\rm Cb}$, and the regions in the green and yellow bands can explain $m_W^{\rm Cb}$ at the $1\sigma$ and $2\sigma$ levels, respectively. Constraints from the oblique parameters are also plotted, which are bound by the blue-dashed contours at the $1\sigma$ confidence level and the red-dashed contours at the $2\sigma$ confidence level. In the left panel, the constraints are implemented in the standard average scenario where $m_t^{\rm Cb} = 171.39 \pm 0.38~{\rm GeV}$ and $m_W^{\rm Cb} = 80.4133 \pm 0.008~{\rm GeV}$ were obtained, while the right panel shows the results for the conservative average scenario where different uncertainties of $m_t$ and $m_W$, i.e., $m_t^{\rm Cb} = 171.39 \pm 1.0~{\rm GeV}$ and $m_W^{\rm Cb} = 80.4133 \pm 0.015~{\rm GeV}$, were adopted~\cite{deBlas:2016ojx}. Bounds from the LHC search for VLQs are represented by the vertical black-dashed line. All of the observables are calculated by the exact formulas listed in Sec. 3 even though the simple analytic expressions in Eqs.~(\ref{U-S}-\ref{U-mW}) are good approximations. Note that there is no $1\sigma$ region for the oblique parameters in the standard average scenario, which is basically because of the specific theoretical structure of the VLQ theory. \label{fig:mw1}}
\end{figure}

To understand the intrinsic physics of the $U$ extension, it is helpful to obtain the leading contributions to the oblique parameters and $m_W$. Up to the $\sin^2 \theta_L^t$ order for the $S$ and $U$ parameters and the $m_t^2 \sin^2 \theta_L^t$  order for the $T$ parameter, these observables are approximated by
\begin{eqnarray}
\text{S} &\simeq & \frac{N_{C}}{6\pi}\left( \frac{2}{3} (s_{L}^t)^2 \ln\frac{m_{T}^{2}}{m_{t}^{2}}-\frac{5}{3} (s_{L}^t)^2 \right),  \quad \text{U} \simeq  \frac{N_{C}}{6\pi} \times \frac{5}{3} \times (s_{L}^t)^2, \label{U-S} \\
\text{T} &\simeq & \frac{N_{C}m_{t}^{2}}{8\pi s_{W}^{2}m_{W}^{2}} \left( (s_{L}^t)^2\ln\frac{m_{T}^{2}}{m_{t}^{2}}- (s_{L}^t)^2 + (s_{L}^t)^4 \frac{m_{T}^2}{2m_{t}^2}  \right), \label{U-T} \\
%\end{eqnarray}
%\begin{eqnarray}
\frac{\delta m_W}{m_W}  & \simeq & \frac{\alpha N_C  }{288 \pi s_W^{2}\left(c_W^2- s_W^{2}\right)} \times \frac{\left(s_{L}^{t}\right)^{2}}{m_W^2}  \times \left \{ \left(18 c_W^2 m_t^{2}- 8 s_W^{2} m_W^{2} \right) \ln \frac{m_T^2}{m_t^2}      \right. \nonumber \\
&& \left.  +10 m_W^{2} - 18  c_W^2 m_t^{2} + 9 \left(s_{L}^{t}\right)^{2} c_W^2  m_T^{2} \right \} \nonumber  \\
&\simeq & 10^{-4} \times \left(\frac{s_{L}^{t}}{0.1}\right)^{2} \times \left \{  1.30 \ln \left (\frac{m_T}{{\rm TeV}}\right )^2 + 3.46 + 0.22 \left(\frac{s_{L}^{t}}{0.1}\right)^{2} \left(\frac{ m_T}{{\rm TeV}}\right)^2 \right \}, \label{U-mW}
\end{eqnarray}
where $N_C=3$ is the color factor, and the latest combined top quark mass, $m_t^{\rm Cb} =171.79~{\rm GeV}$~\cite{deBlas:2016ojx}, is used to obtain the semianalytic approximation of $\delta m_W/m_W$. These formulas reveal the following features:
\begin{itemize}
\item The VLQ contributions arise from the $t^\prime-T^\prime$ mixing, which is reflected by the fact that they are all proportional to $(s_L^t)^2$ at the leading order. Evidently, the larger value the mixing takes, the more significant the effects become.
\item The contributions contain both logarithmic and nonlogarithmic terms. The logarithmic terms reflect the renormalization group running effect in the effective Lagrangian framework, and they are associated with the divergence of the vacuum-polarization diagrams. The nonlogarithmic terms, however, come from the matching between the full theory and the effective theories below the scales of $m_T$ and $m_t$. Computationally speaking, there are two sources for the nonlogarithmic terms in the $U$ extension. One is from the $t$- and $b$-mediated loop contributions subtracted by their corresponding SM predictions.
     The other is from the $W$ self-energy diagram induced by the $W \bar{T} b$ interaction and the $Z$ self-energy diagram induced by the $Z \bar{T} t$ interaction.
\item As indicated by Eq.~(\ref{slt}), $s_L^t \simeq \sqrt{2}\xi_U v/(2 m_T)$, and thus $ (s_L^t)^4 m_T^2 \simeq(s_L^t)^2 \xi_U^2 v^2/2$. It is then inferred that the $(s_L^t)^4 m_T^2/m_t^2$ term in the $T$ parameter is comparable in magnitude to the $(s_L^t)^2$ term if $\xi_U \sim 1$, and therefore, it cannot be neglected in the approximation. Our numerical calculations verify this conclusion. In addition, it is verified that the  $(s_L^t)^4 m_T^2/m_t^2$ term comes from the $Z \bar{T} t$-induced self-energy diagram of $Z$ boson.
\item Generally, the magnitude of the $T$ parameter is much larger than that of the $S$ and $U$ parameters. So $\delta m_W$ is mainly contributed by the $T$ parameter.
\end{itemize}

In Fig.~\ref{fig:mw1}, the capability of the $U$ extension to explain the mass anomaly is studied. Relevant parameter space is projected onto the $m_T-\sin \theta_L^t$ plane.
Constraints from the oblique parameters and the LHC searches for VLQs are also implemented. It is found that the theory can explain the combined experimental value of $m_W$ in broad parameter space. The explanation is consistent with the constraints from the oblique parameters at the $2\sigma$ level in the standard average scenario and at the $1\sigma$ level in the conservative average scenario. In addition, given the relation $\xi_U \simeq \sqrt{2} m_T s_L^t/v \simeq 5.75 \times (m_T/{\rm TeV}) \times s_L^t$ and the results in Fig.~\ref{fig:mw1}, it is inferred $\xi_U \simeq 1.3$ for $m_T = 2~{\rm TeV}$ to predict the central value of $m_W^{\rm Cb}$. Such a $\xi_U$ is consistent with the perturbativity criterion. It is checked that $\delta m_W^T/\delta m_W \simeq 0.996$ for this parameter point, where $\delta m_W^T$ denotes the $T$-term contribution in Eq.~(\ref{m_W}).

\begin{figure}[t]	
	\centering
	\includegraphics[scale=0.33]{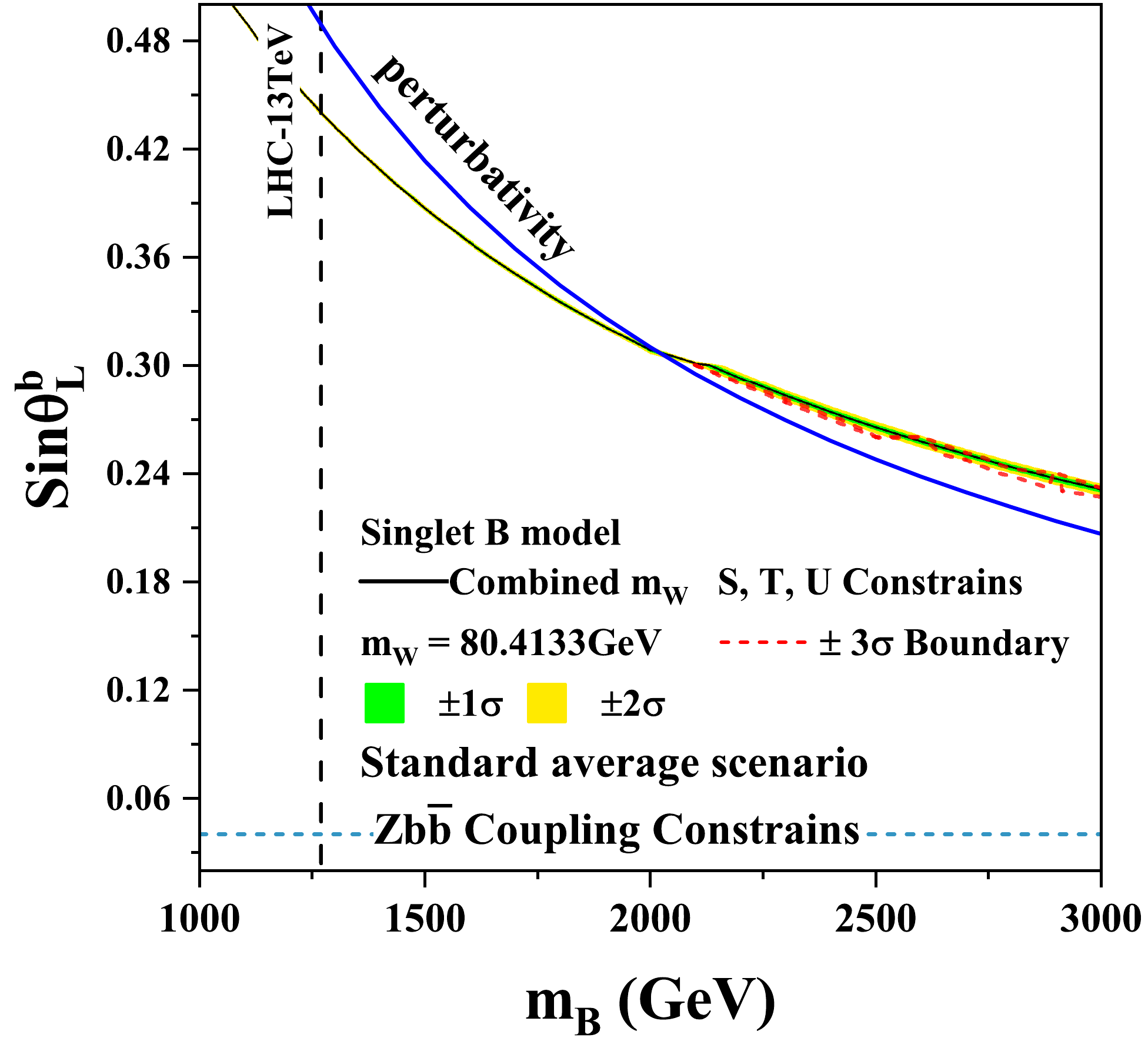}
	\includegraphics[scale=0.33]{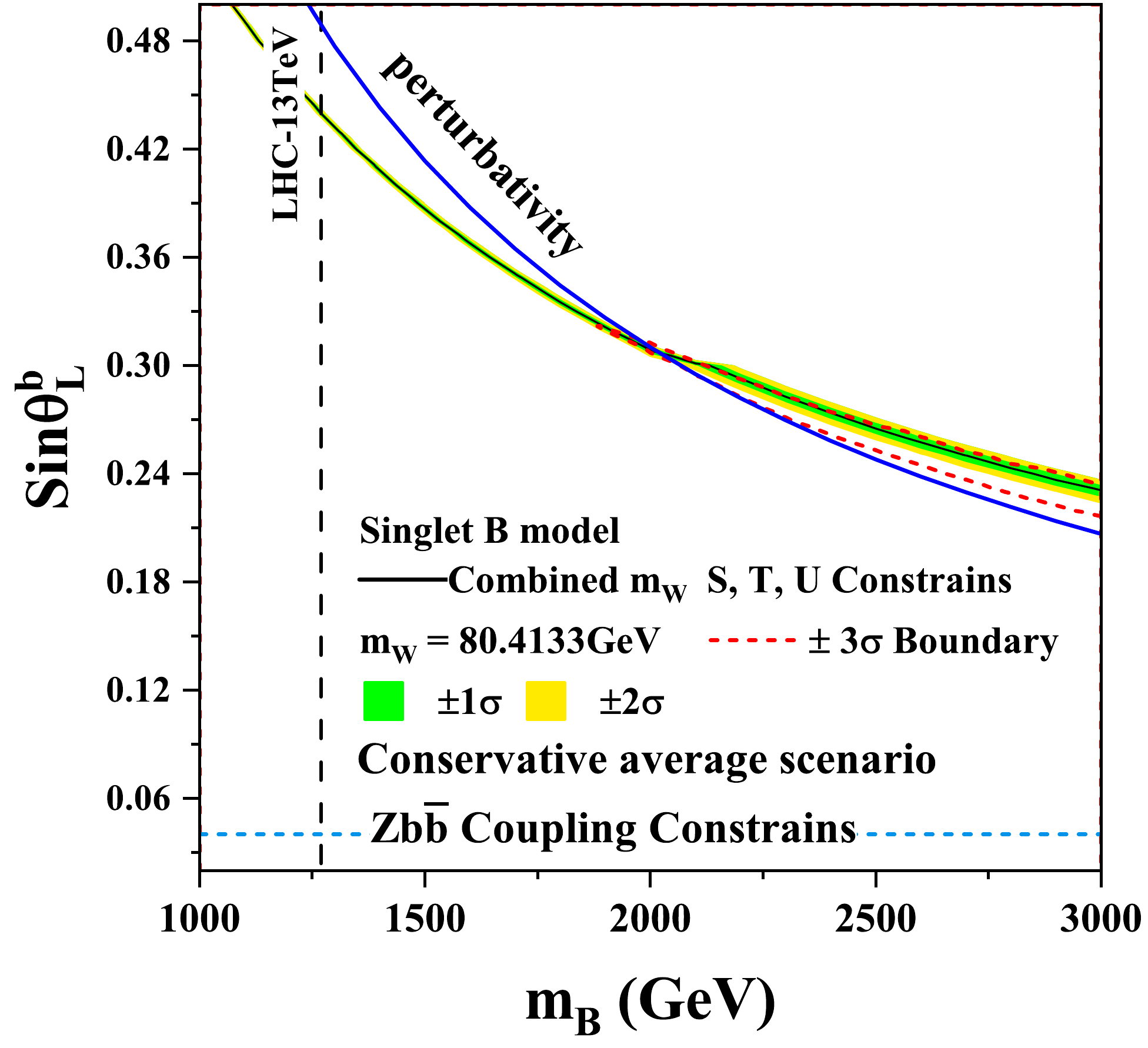}

\vspace{-0.4cm}

	\caption{Similar to Fig.\ref{fig:mw1}, but for the $D$ extension with the additional $Z b \bar{b}$ coupling constraints imposing an upper bound on $\sin \theta_L^b$, denoted by the blue-dashed lines. Since explaining the $W$-boson mass anomaly requires a relatively large $\sin \theta_L^b$, constraints from the perturbativity are also plotted, which are labeled by the blue curves. In addition, given that the extension is hardly consistent with the constraints from the oblique parameters at the $2\sigma$ level, only the $3 \sigma$ boundaries of the parameters are presented as red-dashed lines in this figure.   \label{fig:mw2}}
\end{figure}

\subsection{$D$ extension}

Similar to the $U$ extension, the oblique parameters and $m_W$ in the $D$ extension are approximated by
\begin{eqnarray}
\text{S} &\simeq& \frac{N_{C}}{6\pi} \left \{ \frac{4}{3} (s_{L}^{b})^{2}  \ln \frac{m_B^2}{m_b^2}  -  \frac{5}{3} (s_{L}^{b})^{2} \right \} , \\
\text{T} &\simeq & \frac{N_{C}m_{t}^{2}}{8\pi s_{W}^{2}m_{W}^{2}} \left( -(s_{L}^{b})^{2}\ln\frac{m_{B}^{2}}{m_{t}^{2}}+ (s_{L}^{b})^{4} \frac{m_{B}^2}{2m_{t}^2} \right), \\
\text{U} &\simeq& \frac{N_{C}}{6\pi} \left \{ -2 (s_{L}^{b})^{2}  \ln \frac{m_t^2}{m_b^2}  +  \frac{5}{3} (s_{L}^{b})^{2} \right \} , \\
\frac{\delta m_W}{m_W}  &\simeq &  - \frac{\alpha N_C  }{288 \pi s_W^{2}\left(c_W^2- s_W^{2}\right)} \times \frac{\left(s_{L}^{b}\right)^{2}}{m_W^2} \times \left \{ \left[ 12\left(c_W^2-s_W^2\right)m_W^2 - 18 c_W^2 m_t^2 \right] \ln \frac{m_t^2}{m_b^2} \right.  \nonumber \\
& &\left.  + \left(16 s_W^2 m_W^2 + 18 c_W^2 m_t^2\right) \ln \frac{m_B^2}{m_b^2} -10 m_W^{2} - 9 \left(s_{L}^{b}\right)^{2} c_W^2 m_B^{2} \right \}  \nonumber  \\
& \simeq & - 10^{-4} \times  \left(\frac{s_{L}^{b}}{0.1}\right)^{2}\times \left \{ 1.42 \ln \left ( \frac{m_B}{{\rm TeV}}\right)^2 + 6.36 - 0.23 \left(\frac{s_{L}^{b}}{0.1}\right)^{2} \left(\frac{ m_B}{{\rm TeV}}\right)^2 \right \} \nonumber \\
& \simeq & - 10^{-2} \times \frac{\xi_D^2}{4 \pi} \times \left ( \frac{{\rm TeV}}{m_B} \right )^2 \times \left \{ 0.54 \ln \left ( \frac{m_B}{{\rm TeV}}\right)^2 + 2.42 - 3.25 \times \frac{\xi_D^2}{4\pi} \right \}.
\end{eqnarray}
In getting the last approximation of $\delta m_W/m_W$, the relation $\sin \theta_L^b \simeq \sqrt{2} \xi_D v/(2 m_B)$ is used, and it shows that the mass correction decreases as the state $B$ becomes heavy for fixed $\xi_D$.

These approximations reveal the following facts:
\begin{itemize}
\item For $s_L^b \lesssim 0.04$ required by the $Z b \bar{b}$ coupling measurements~\cite{Aguilar-Saavedra:2013qpa}, the mass correction is negative for $m_B \lesssim 20~{\rm TeV}$, which can be learned by the first semianalytic form.
\item For $m_B \gtrsim 20~{\rm TeV}$ and $\xi_D^2 \leq 4 \pi$, the correction remains negative in the second semianalytic form.
\end{itemize}
They reflect that the extension cannot explain the $W$-boson mass anomaly on the premise of satisfying all the constraints.

In the following, the implications of the mass anomaly on the $D$ extension are studied numerically. In Fig.~\ref{fig:mw2}, the prediction of $m_W$ in the $D$ extension along with various constraints are presented. It shows that, although the extension can explain the anomaly in very narrow parameter space, the explanation always conflicts with the constraints from $Z b \bar{b}$ couplings. It also conflicts with the oblique parameters for $m_B \lesssim 2~{\rm TeV}$ and the perturbativity for $m_B \gtrsim 2~{\rm TeV}$.  These features are consistent with the analyses of the approximations.

\newpage

\begin{figure}[t]	
	\centering
	\includegraphics[scale=0.33]{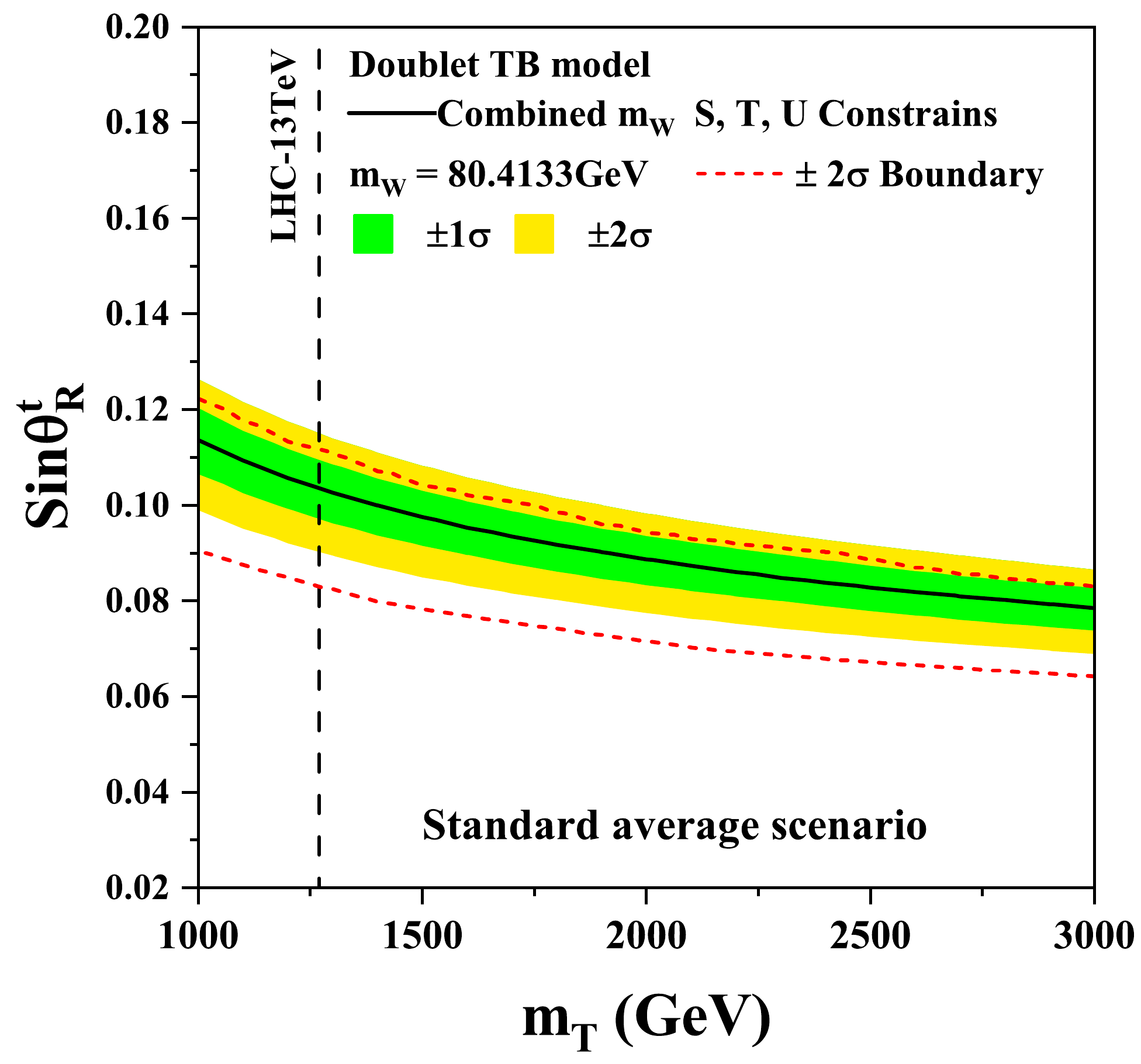}
	\includegraphics[scale=0.33]{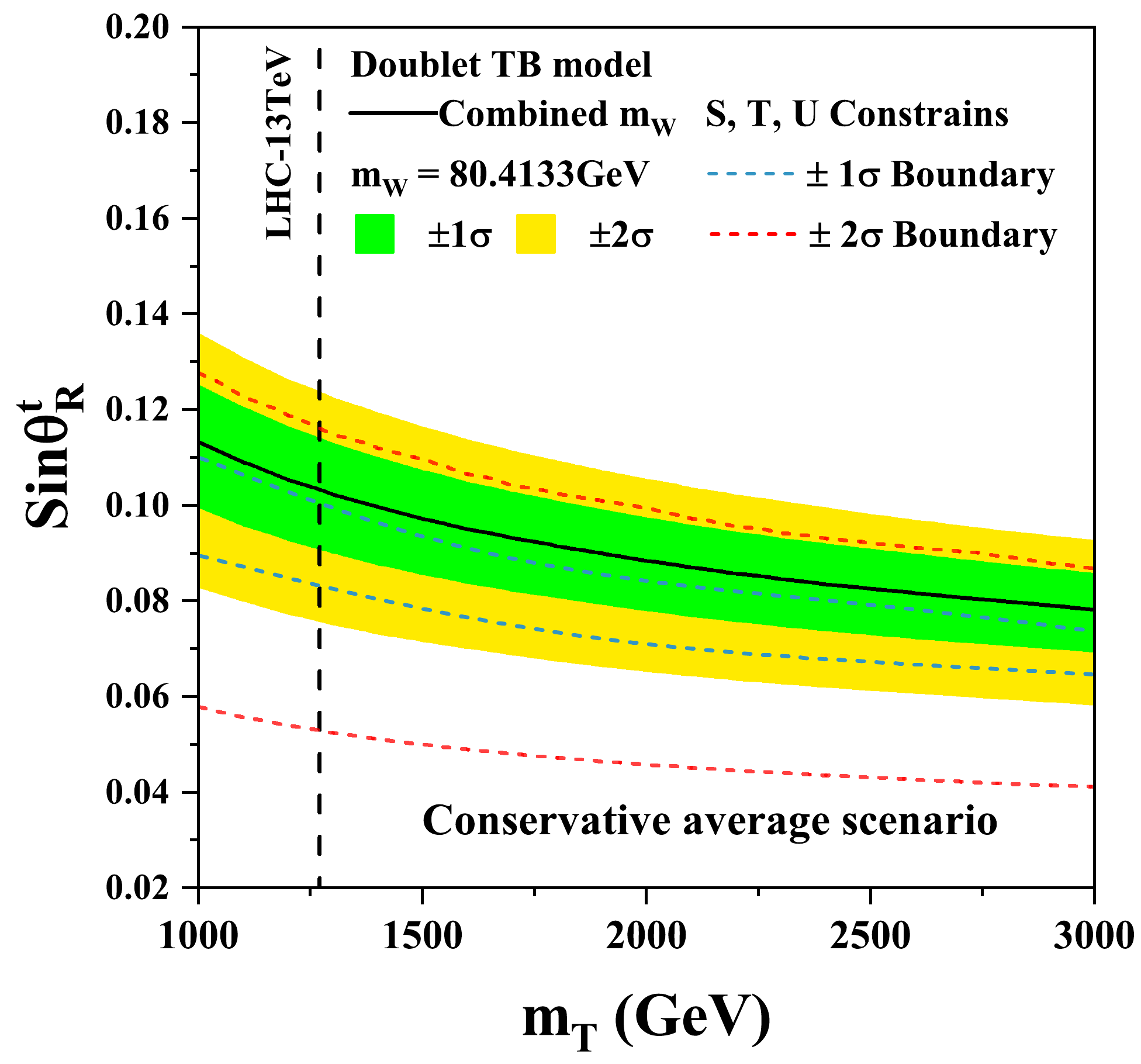}

\vspace{-0.4cm}

	\caption{Similar to Fig.\ref{fig:mw1} but for the $Q_1$ extension.\label{fig:mw3}}
\end{figure}

\subsection{$Q_1$ extension}

The oblique parameters and $m_W$ in the $Q_1$ extension are approximated by
\begin{eqnarray}
\text{S} &\simeq & \frac{N_{C}}{6\pi} \left( \frac{4}{3}(s_{R}^{t})^{2}\ln\frac{m_{T}^{2}}{m_{t}^{2}}-\frac{7}{3}(s_{R}^{t})^{2} \right), \quad \text{U} \simeq  \frac{N_{C}}{6\pi} (s_{R}^{t})^{2},   \\
\text{T} &\simeq & \frac{N_{C}m_{t}^{2}}{8\pi s_{W}^{2}m_{W}^{2}} \left( 2(s_{R}^{t})^{2}\ln\frac{m_{T}^{2}}{m_{t}^{2}}-3(s_{R}^{t})^{2}+ (s_{R}^{t})^{4} \frac{2m_{T}^2}{3m_{t}^2} \right),   \\
\frac{\delta m_W}{m_W}    & \simeq & \frac{\alpha N_C  }{288 \pi s_W^{2}\left(c_W^2- s_W^{2}\right)} \times \frac{\left(s_{R}^{t}\right)^{2}}{m_W^2}   \times \left \{ \left(36 c_W^2 m_t^{2} - 16 s_W^{2} m_W^{2} \right) \ln \frac{m_T^2}{m_t^2}   \right.  \nonumber \\
&& \left. +  \left( 6 + 16 s_W^2 \right) m_W^{2} -54 c_W^2 m_t^{2} + 12 \left(s_{R}^{t}\right)^{2} c_W^2 m_T^{2} \right \}  \nonumber  \\
& \simeq & 10^{-4} \times \left(\frac{s_{R}^{t}}{0.1}\right)^{2}\times  \left \{ 2.60 \ln \left (\frac{m_T}{{\rm TeV}}\right)^2 + 5.35 + 0.31 \left(\frac{s_{R}^{t}}{0.1}\right)^{2} \left(\frac{ m_T}{{\rm TeV}}\right)^2 \right \}.
\end{eqnarray}

In Fig.~\ref{fig:mw3}, the $W$-boson mass anomaly is investigated, and roughly the same conclusions as those in Fig.~\ref{fig:mw1} are drawn. A slight difference comes from the fact that this model prefers a smaller mixing angle than the $U$ extension, such as $s_R^t \simeq 0.095$ (or equivalently $\xi_{Q_1} \simeq 1.1$), for $m_T = 2~{\rm TeV}$ to predict the central value of $m_W^{\rm CDF}$. In addition, it is checked that $\delta m_W^T/\delta m_W \simeq 1.02$ for this parameter point.

\newpage

\begin{figure}[t]	
	\centering
	\includegraphics[scale=0.33]{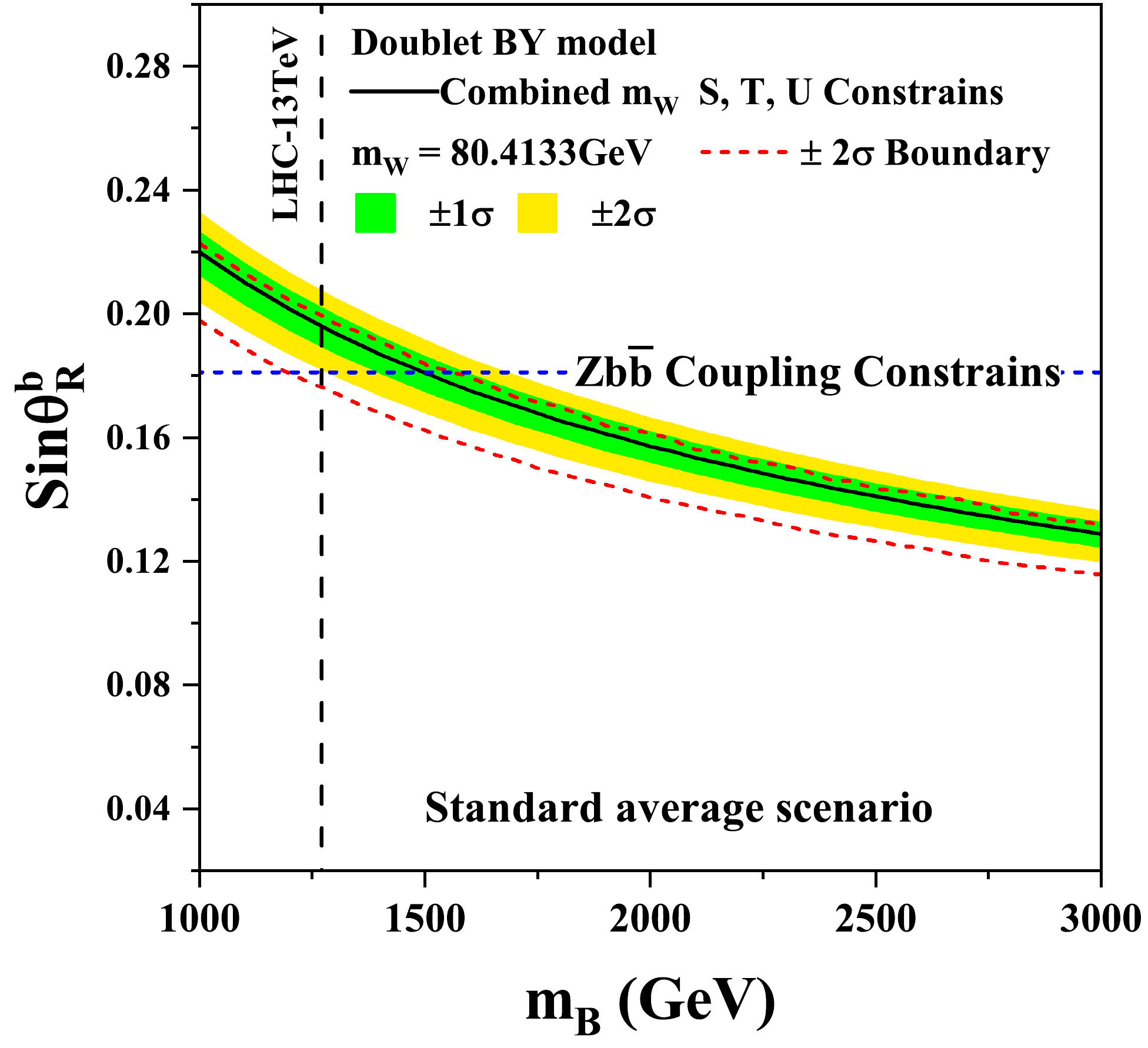}
	\includegraphics[scale=0.33]{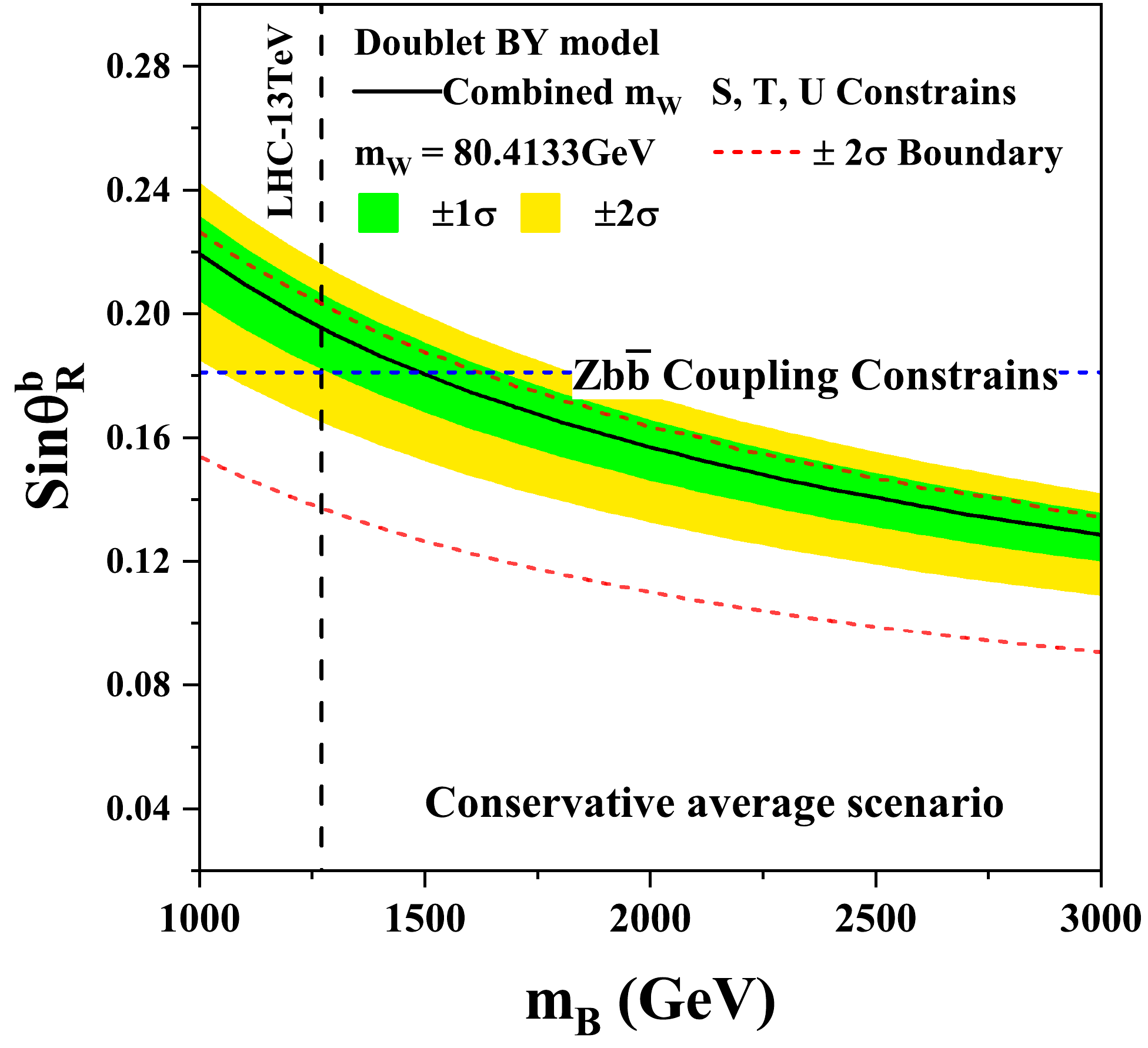}

\vspace{-0.4cm}

	\caption{Similar to Fig.\ref{fig:mw1} but for the $Q_5$ extension. One new feature is that the $Z b \bar{b}$ couplings can impose an upper bound on $\sin \theta_R^b$ (the blue-dashed line), which is weaker than the oblique parameters in limiting the theory for $m_B \gtrsim 1.6~{\rm TeV}$ in the standard average scenario. Note that compared with the predictions of the $U$ and $Q_1$ extensions, the $2 \sigma$ region for the oblique parameters in the standard average scenario is narrowed, and the $1\sigma$ region in the conservative average scenario is absent. These behaviors reflect that the $Q_5$ extension is less compatible with the latest EWPOs. \label{fig:mw4}}
\end{figure}

%\vspace{-0.5cm}

\subsection{$Q_5$ extension}

%\vspace{-0.3cm}

The oblique parameters and $m_W$ in the $Q_5$ extension are approximated by
\begin{eqnarray}
\text{S} &\simeq& \frac{N_{C}}{6\pi} \left \{- \frac{2}{3} (s_{R}^{b})^{2}  \ln \frac{m_B^2}{m_b^2}  +  \frac{11}{3} (s_{R}^{b})^{2} \right \} ,\quad \text{U} \simeq - \frac{N_{C}}{6\pi}  (s_{R}^{b})^{2} , \\
\text{T} &\simeq & \frac{N_{C}m_{t}^{2}}{8\pi s_{W}^{2}m_{W}^{2}}  (s_{R}^{b})^{4}  \frac{2m_{B}^2}{3m_{t}^2} , \\
\frac{\delta m_W}{m_W}  &\simeq &   \frac{\alpha N_C  }{288 \pi s_W^{2}\left(c_W^2- s_W^{2}\right)} \times \frac{\left(s_{R}^{b}\right)^{2}}{m_W^2} \times \left \{ 8 s_W^2  m_W^2 \ln \frac{m_B^2}{m_b^2} \right.  \nonumber \\
& &\left.  - \left(6+32s_W^2\right) m_W^{2}  + 12 \left(s_{R}^{b}\right)^{2} c_W^2 m_B^{2} \right \}  \nonumber  \\
& \simeq & 10^{-4} \times \left(\frac{s_{R}^{b}}{0.1}\right)^{2}\times  \left \{ 0.04 \ln \left ( \frac{m_B}{{\rm TeV}}\right)^2 + 0.13 + 0.31 \left(\frac{s_{R}^{b}}{0.1}\right)^{2} \left(\frac{ m_B}{{\rm TeV}}\right)^2 \right \}.
\end{eqnarray}
Similar to the $U$ extension, its capability to interpret the anomaly is studied in Fig.~\ref{fig:mw4}. This figure shows that $\sin \theta_R^b \simeq 0.165$, or equivalently $\xi_{Q_5} \simeq 1.9$, for $m_B = 2~{\rm TeV}$ is needed to predict the central value of $m_W^{\rm Cb}$. In this case, $\delta m_W^T/\delta m_W \simeq 0.93$. At this stage, it should be clarified that $\xi_{Q_5}$ parametrizes the coupling of the $b_R^\prime$ field to VLQ multiplets, and $\xi_{Q_5} > y_{t^\prime} \gg Y_{b^\prime}$ looks unnatural. In fact, the possibility that $b_R^\prime$ couples strongly to new physics lacks solid theoretical motivations in model building~\cite{Cvetic:1997eb}.

%\vspace{-0.5cm}
\newpage

\begin{figure}[t]	
	\centering
	\includegraphics[scale=0.33]{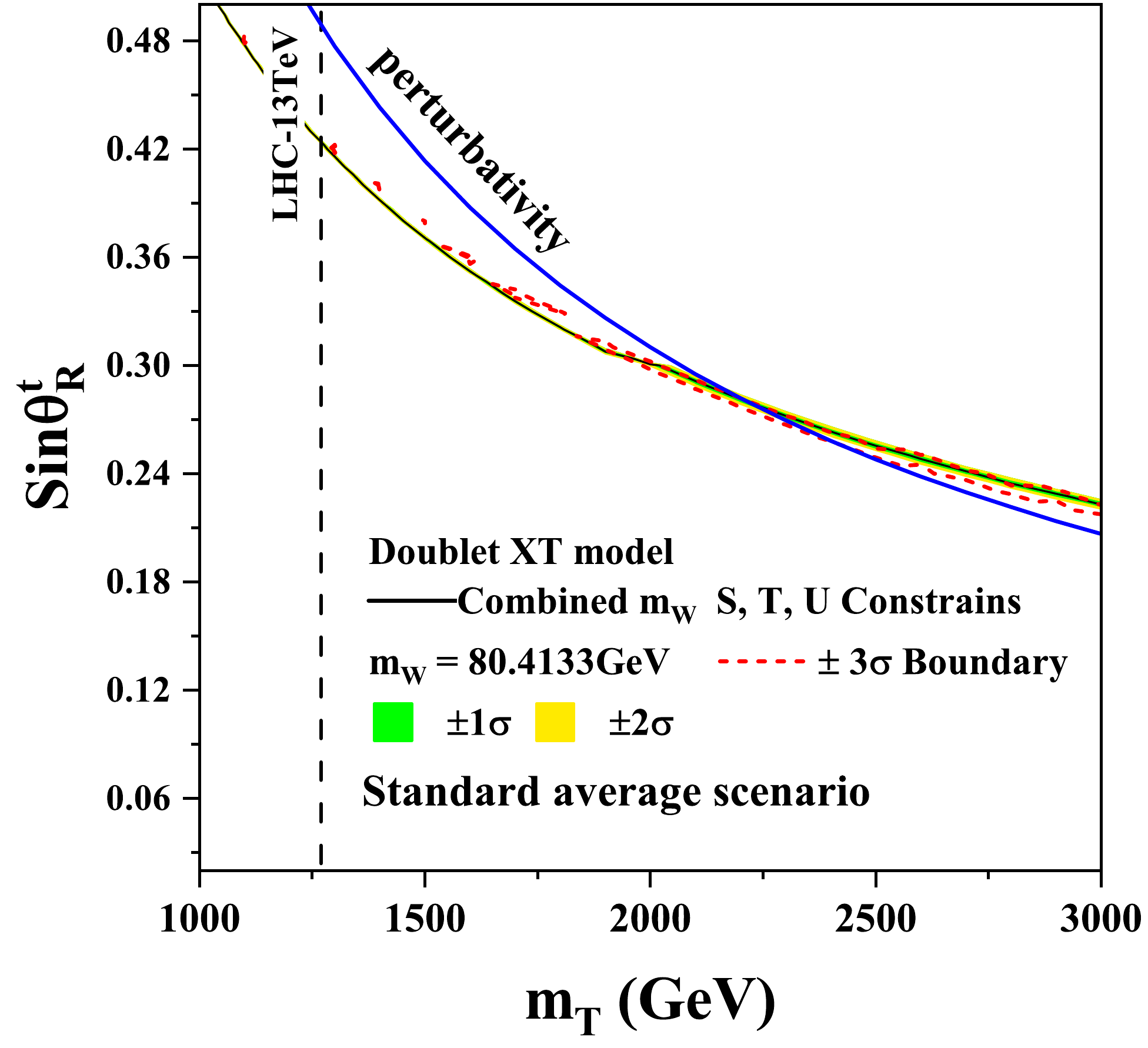}
	\includegraphics[scale=0.33]{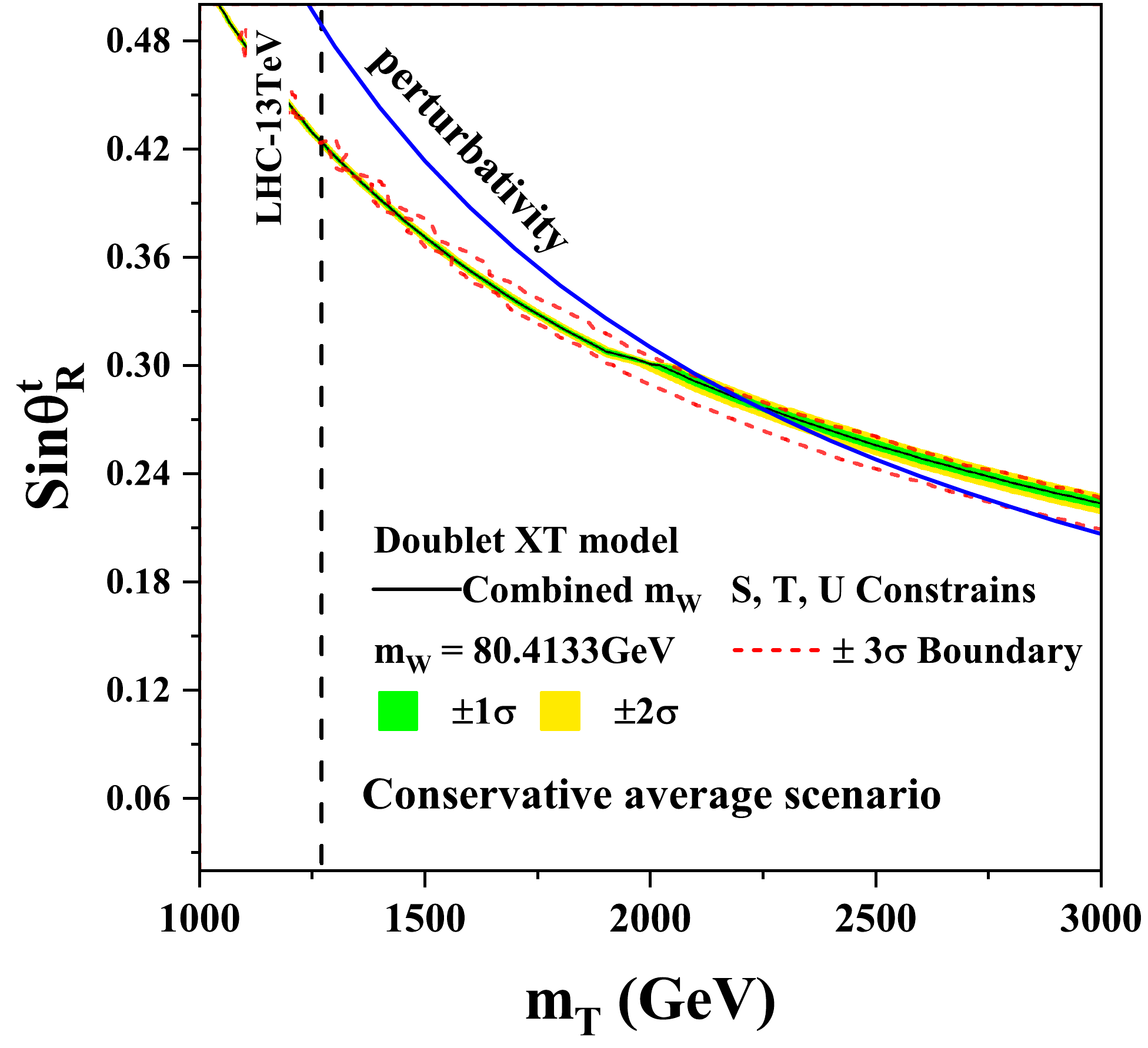}

\vspace{-0.4cm}

	\caption{Similar to Fig.\ref{fig:mw2} but for the $Q_7$ extension, the $Z b \bar{b}$ couplings are unable to set tight constraints due to the absence of the tree-level $b^\prime-B^\prime$ mixing. \label{fig:mw5}}
\end{figure}

\subsection{$Q_7$ extension}

%\vspace{-0.2cm}

The oblique parameters and $m_W$ in the $Q_7$ extension are approximated by
\begin{eqnarray}
\text{S} &\simeq & \frac{N_{C}}{6\pi} \left( -\frac{4}{3} (s_{R}^t)^2 \ln\frac{m_{T}^{2}}{m_{t}^{2}}+ 5 (s_{R}^t)^2  \right), \quad \text{U} \simeq  -\frac{N_{C}}{6\pi} (s_{R}^t)^2,  \\
\text{T} &\simeq & \frac{N_{C}m_{t}^{2}}{8\pi s_{W}^{2}m_{W}^{2}} \left( -2 (s_{R}^t)^2 \ln\frac{m_{T}^{2}}{m_{t}^{2}}+ 3(s_{R}^t)^2 + (s_{R}^t)^4 \frac{2m_{T}^2}{3m_{t}^2} \right),  \\
\frac{\delta m_W}{m_W}  & \simeq & - \frac{\alpha N_C  }{288 \pi s_W^{2}\left(c_W^2- s_W^{2}\right)} \times \frac{\left(s_{R}^{t}\right)^{2}}{m_W^2} \times \left \{ \left(36  c_W^2 m_t^{2}- 16 s_W^{2} m_W^{2} \right)\ln \frac{m_T^2}{m_t^2}   \right. \nonumber \\
& &\left. + \left(6 + 48 s_W^2 \right) m_W^{2} - 54 c_W^2 m_t^{2} - 12  \left(s_{R}^{t}\right)^{2} c_W^2  m_T^{2} \right \}  \nonumber \\
& \simeq & - 10^{-4} \times \left(\frac{s_{R}^{t}}{0.1}\right)^{2}\times \left \{ 2.61 \ln \left ( \frac{m_T}{{\rm TeV}}\right )^2 + 5.50 - 0.31 \left(\frac{s_{R}^{t}}{0.1}\right)^{2} \left(\frac{ m_T}{{\rm TeV}}\right)^2 \right \} \nonumber \\
& \simeq & - 10^{-2} \times \frac{\xi_{Q_7}^2}{4 \pi} \times \left ( \frac{{\rm TeV}}{m_T} \right )^2 \times \left \{ 0.99 \ln \left ( \frac{m_T}{{\rm TeV}}\right)^2 + 2.09 - 4.40 \times \frac{\xi_{Q_7}^2}{4\pi} \right \}.
\end{eqnarray}
Similar to the analyses of the $D$ extension, the correction to $m_W$ is always negative for $s_R^t \lesssim 0.1$, and the $W$-boson mass anomaly can be explained only in very narrow parameter space characterized by a large $\sin \theta_R^t$. These features are shown in Fig.~{\ref{fig:mw5} where the prediction of $m_W$ as well as different constraints are presented. It should be pointed out that, although the explanation for $m_T \lesssim 2~{\rm GeV}$ is consistent with all the constraints, it is disfavored by the vacuum stability discussed at the end of this section.

%\vspace{-0.5cm}

\newpage

\begin{figure}[t]	
	\centering
	\includegraphics[scale=0.33]{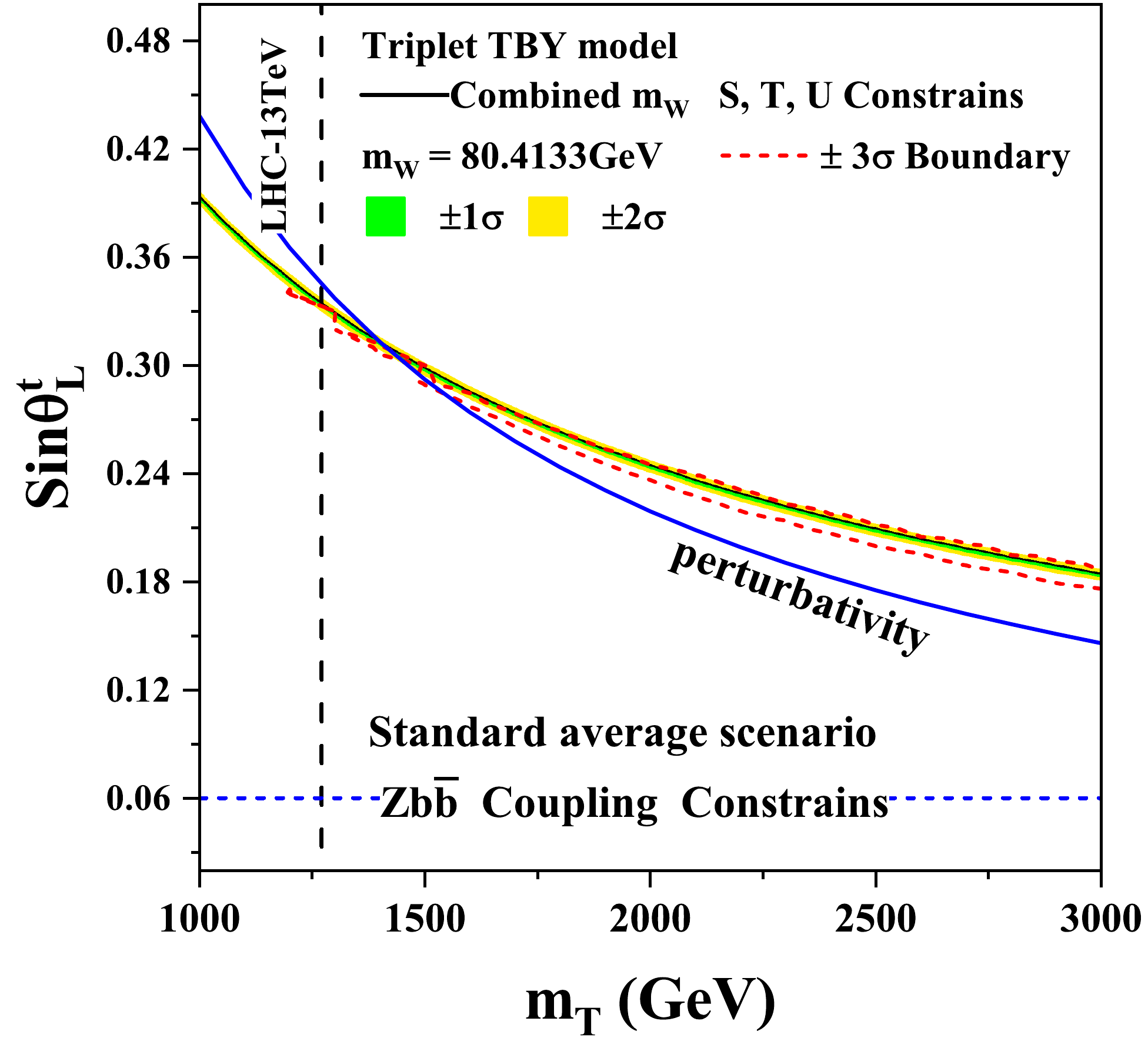}
	\includegraphics[scale=0.33]{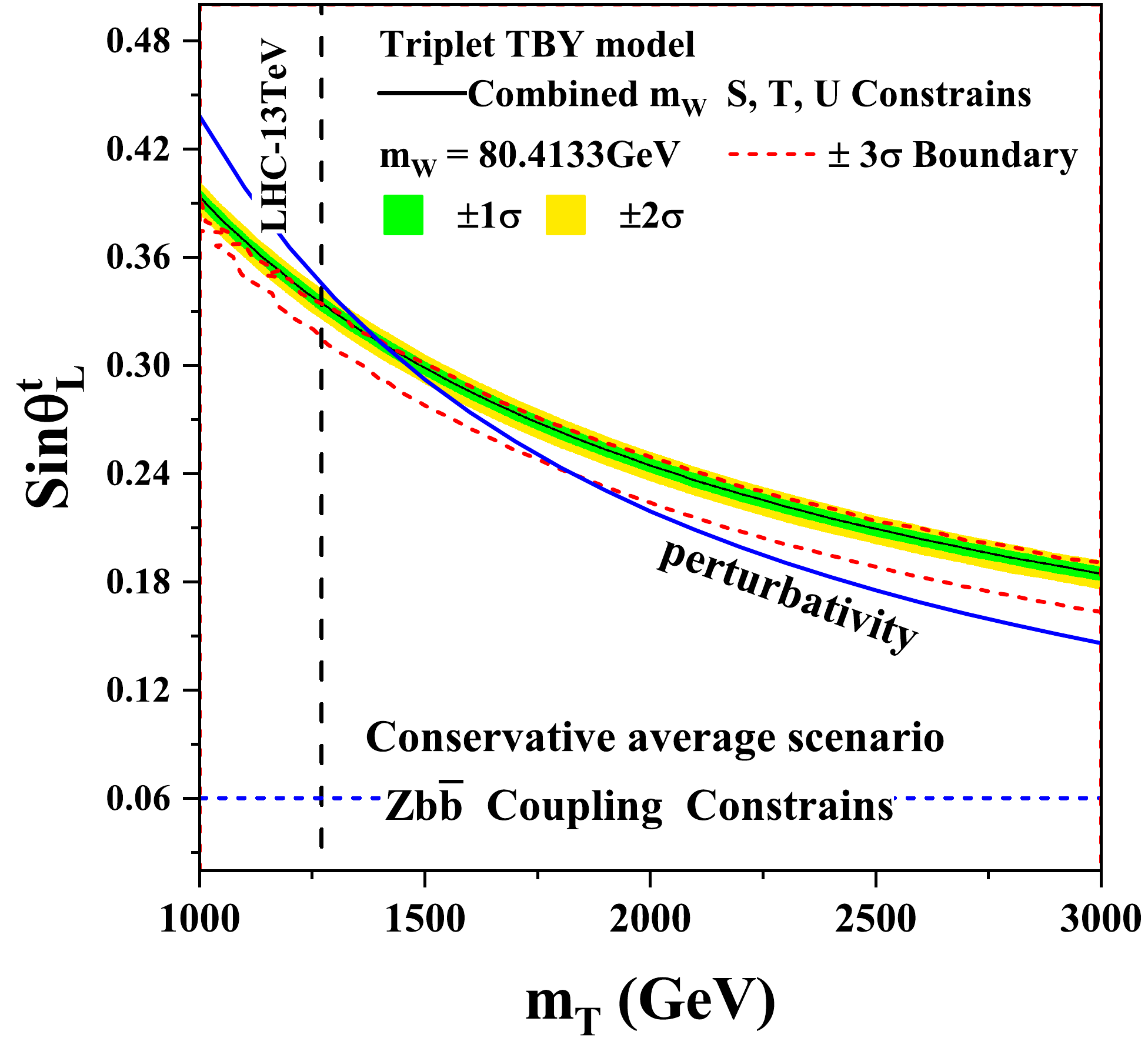}

\vspace{-0.4cm}

	\caption{Similar to Fig.\ref{fig:mw2} but for the $T_1$ extension, the perturbativity is tighter in limiting the $m_T-\sin \theta_L^t$ plane. \label{fig:mw6}}
\end{figure}

\subsection{$T_1$ extension}

%\vspace{-0.2cm}

The oblique parameters and $m_W$ in the $T_1$ extension are approximated by
\begin{eqnarray}
\text{S} &\simeq& \frac{N_{C}}{6\pi} \left \{ \frac{2}{3} (s_{L}^{t})^{2} \left ( \ln \frac{m_B^2}{m_b^2} - \ln \frac{m_T^2}{m_t^2}  \right ) +  \frac{13}{6} (s_{L}^{t})^{2} \right \} \nonumber  \\
&\simeq & \frac{N_{C}}{6\pi} \left( \frac{2}{3}(s_{L}^{t})^{2}\ln\frac{m_{t}^{2}}{m_{b}^{2}}+\frac{13}{6}(s_{L}^{t})^{2} \right), \\
\text{T} &\simeq & \frac{N_{C}m_{t}^{2}}{8\pi s_{W}^{2}m_{W}^{2}} \left( -\frac{3}{2}(s_{L}^{t})^{2}\ln\frac{m_{T}^{2}}{m_{t}^{2}}+3(s_{L}^{t})^{2}+ (s_{L}^{t})^{4} \frac{19m_{T}^2}{24m_{t}^2} \right), \\
\text{U} &\simeq& \frac{N_{C}}{6\pi} \left \{ - (s_{L}^{t})^{2} \left ( \ln \frac{m_B^2}{m_b^2} - \ln \frac{m_T^2}{m_t^2}  \right ) -  \frac{5}{6} (s_{L}^{t})^{2} \right \} \nonumber  \\
&\simeq & \frac{N_{C}}{6\pi} \left( -(s_{L}^{t})^{2}\ln\frac{m_{t}^{2}}{m_{b}^{2}}-\frac{5}{6}(s_{L}^{t})^{2} \right), \\
\frac{\delta m_W}{m_W}  &\simeq &  - \frac{\alpha N_C  }{288 \pi s_W^{2}\left(c_W^2- s_W^{2}\right)} \times \frac{\left(s_{L}^{t}\right)^{2}}{m_W^2} \times \left \{ \left( 6 - 4 s_W^2\right) m_W^2 \ln \frac{m_t^2}{m_b^2} \right.  \nonumber \\
& &\left.  + 27 c_W^2  m_t^{2} \ln \frac{m_T^2}{m_t^2} + \left(5+16s_W^2\right) m_W^{2} - c_W^2 54 m_t^{2} - \frac{ 57}{4} \left(s_{L}^{t}\right)^{2} c_W^2 m_T^{2} \right \}  \nonumber  \\
& \simeq & -  10^{-4} \times \left(\frac{s_{L}^{t}}{0.1}\right)^{2}\times \left \{ 2.01 \ln \left ( \frac{m_T}{{\rm TeV}}\right)^2 + 4.05 - 0.36 \left(\frac{s_{L}^{t}}{0.1}\right)^{2} \left(\frac{ m_T}{{\rm TeV}}\right)^2 \right \} \nonumber \\
& \simeq & - 10^{-2} \times \frac{\xi_{T_1}^2}{4 \pi} \times \left ( \frac{{\rm TeV}}{m_T} \right )^2 \times \left \{ 0.77 \ln \left ( \frac{m_T}{{\rm TeV}}\right)^2 + 1.53 - 5.20 \times \frac{\xi_{T_1}^2}{4\pi} \right \}.
\end{eqnarray}
Given that the $Z b \bar{b}$ coupling constraints require $s_L^t < 0.06 $~\cite{Aguilar-Saavedra:2013qpa}, the mass correction is always negative, which can be obtained in the same way as the discussion of the $D$ extension. Numerical results are presented in  Fig.~\ref{fig:mw6}, which are similar to those of Fig.~\ref{fig:mw2} except that the perturbativity constraint becomes tighter.

% \newpage

\begin{figure}[t]	
	\centering
	\includegraphics[scale=0.33]{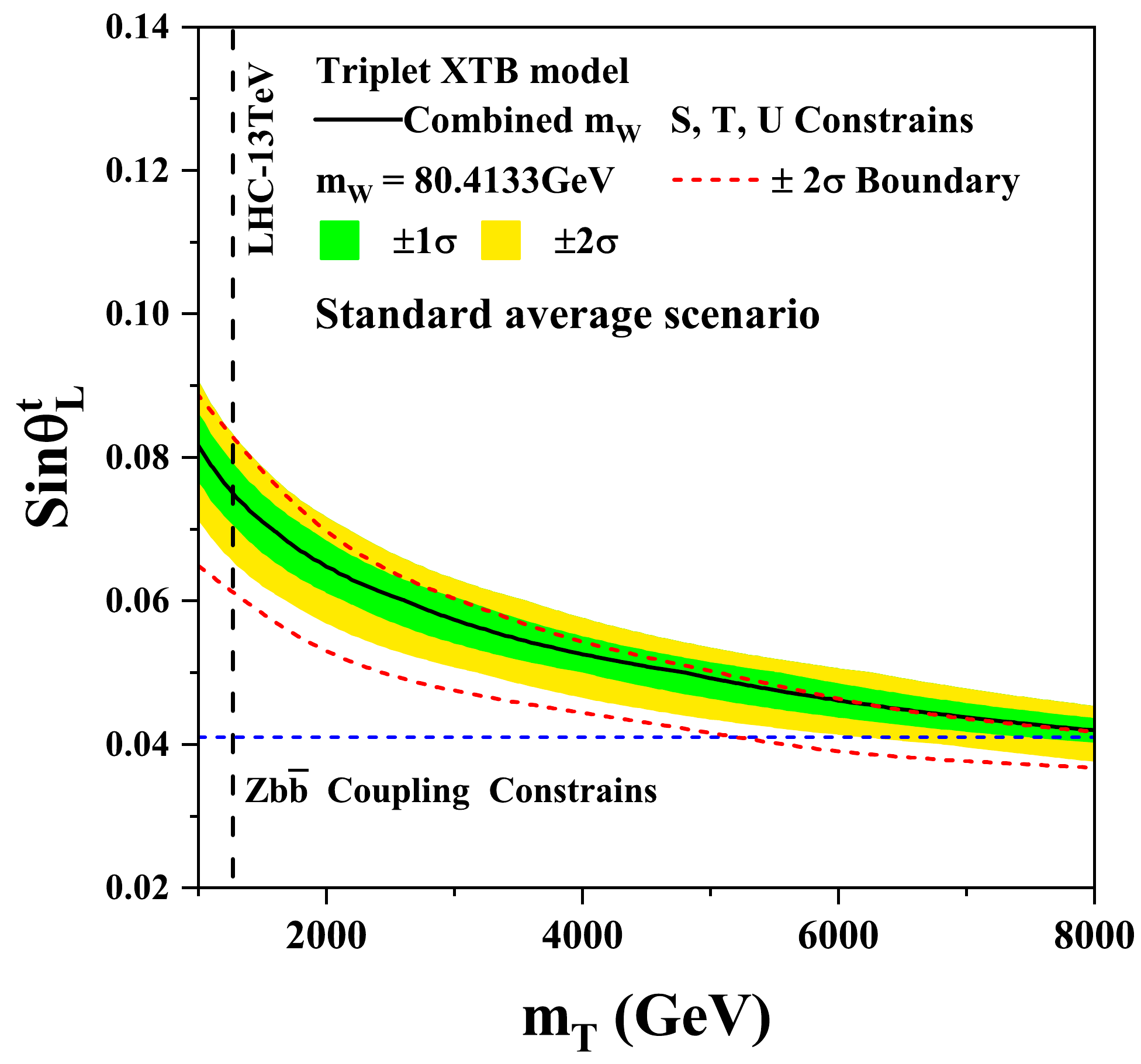}
	\includegraphics[scale=0.33]{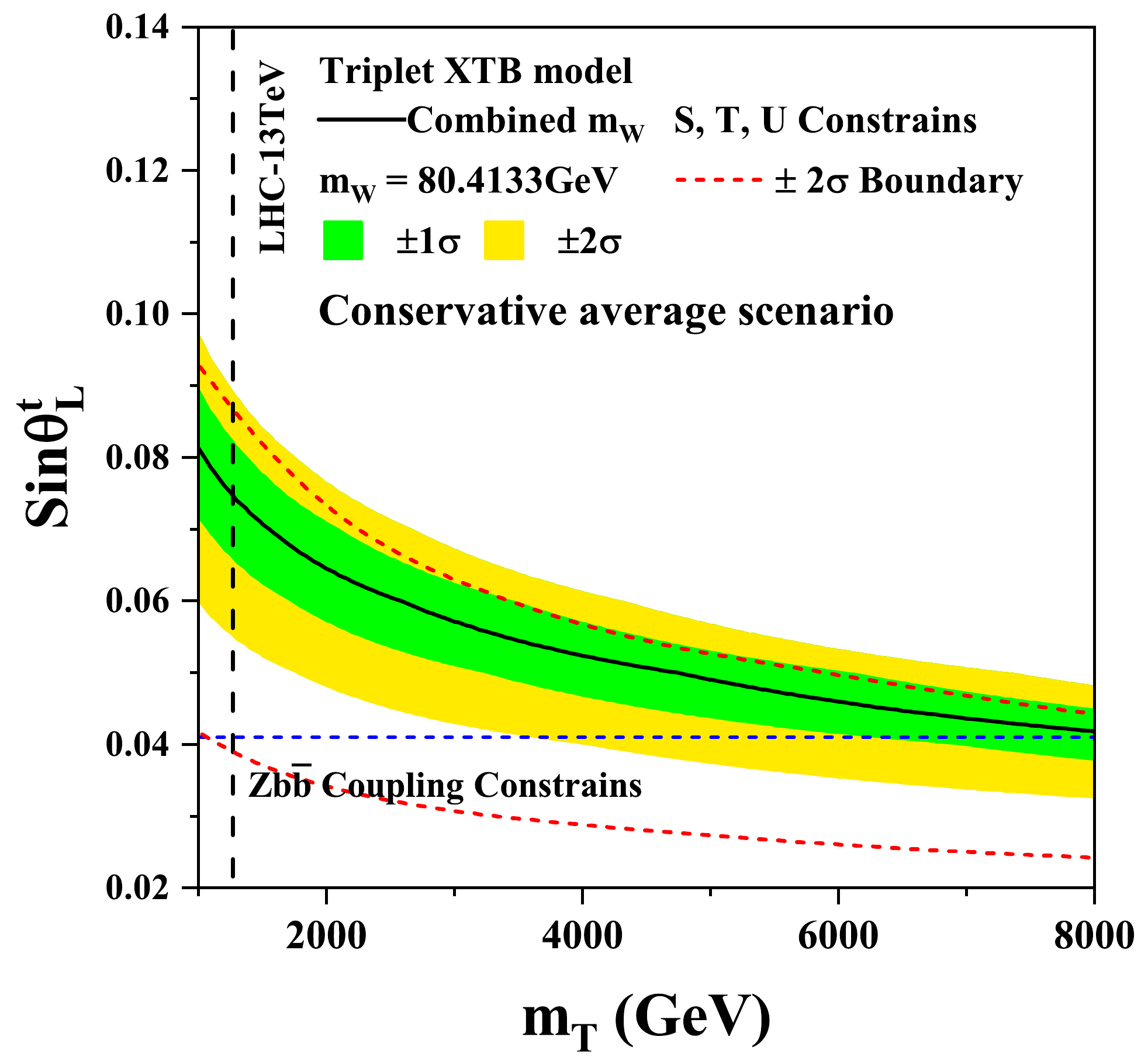}

\vspace{-0.4cm}

	\caption{Similar to Fig.\ref{fig:mw4} but for the $T_2$ extension, this figure indicates that the oblique parameters and the $Z b \bar{b}$ couplings prefer different parameter regions for $m_T \lesssim 5~{\rm TeV}$ in the standard average scenario, so the range of $m_T$ is extended up to $8~{\rm TeV}$ in studying the mass anomaly. \label{fig:mw7}}
\end{figure}

%\vspace{-0.5cm}

\subsection{$T_2$ extension}

%\vspace{-0.3cm}

The oblique parameters and $m_W$ in this model are approximated by
\begin{eqnarray}
\text{S} &\simeq& \frac{N_{C}}{6\pi} \left \{ \frac{16}{3} \left ( \ln \frac{m_B^2}{m_b^2} - \ln \frac{m_t^2}{m_b^2} - \ln \frac{m_T^2}{m_t^2}  \right ) -
\frac{8}{3} (s_{L}^{t})^{2} \ln \frac{m_B^2}{m_b^2} + \frac{2}{3} (s_{L}^{t})^{2} \ln \frac{m_T^2}{m_t^2} +  \frac{13}{3} (s_{L}^{t})^{2} \right \} \nonumber  \\
&\simeq & \frac{N_{C}}{6\pi} \left( -\frac{8}{3}(s_{L}^{t})^{2}\ln\frac{m_{t}^{2}}{m_{b}^{2}}-2(s_{L}^{t})^{2}\ln\frac{m_{T}^{2}}{m_{t}^{2}}+\frac{29}{3}(s_{L}^{t})^{2} \right), \\
\text{T} &\simeq & \frac{N_{C}m_{t}^{2}}{8\pi s_{W}^{2}m_{W}^{2}} \left( 3(s_{L}^{t})^{2}\ln\frac{m_{T}^{2}}{m_{t}^{2}}-5(s_{L}^{t})^{2}+ (s_{L}^{t})^{4} \frac{19m_{T}^2}{6m_{t}^2} \right),   \\
\text{U}  &\simeq& \frac{N_{C}}{6\pi} \left \{ 4 \left ( \ln \frac{m_T^2}{m_t^2}  + \ln \frac{m_t^2}{m_b^2} - \ln \frac{m_B^2}{m_b^2} \right ) +
2 (s_{L}^{t})^{2} \left ( \ln \frac{m_B^2}{m_b^2} + \ln \frac{m_t^2}{m_b^2} - \ln \frac{m_T^2}{m_t^2} \right ) +  \frac{7}{3} (s_{L}^{t})^{2} \right \} \nonumber  \\
&\simeq & \frac{N_{C}}{6\pi} \left( 4(s_{L}^{t})^{2}\ln\frac{m_{t}^{2}}{m_{b}^{2}}-\frac{5}{3}(s_{L}^{t})^{2} \right), \\
\frac{\delta m_W}{m_W} & \simeq & \frac{\alpha N_C  }{288 \pi s_W^{2}\left(c_W^2- s_W^{2}\right)} \times \frac{\left(s_{L}^{t}\right)^{2}}{m_W^2} \times \left \{ \left( 24 - 16 s_W^2\right) m_W^2 \ln \frac{m_t^2}{m_b^2} \right.  \nonumber  \\
& &\left. + \left( 24 s_W^{2} m_W^{2} + 54 c_W^2 m_t^{2} \right) \ln \frac{m_T^2}{m_t^2} -\left(10+96s_W^2\right) m_W^{2}  -90  c_W^2 m_t^{2} + 57 \left(s_{L}^{t}\right)^{2} c_W^2 m_T^{2} \right \} \nonumber  \\
& \simeq &  10^{-4} \times \left(\frac{s_{L}^{t}}{0.1}\right)^{2}\times \left \{ 4.13 \ln \left (\frac{m_T}{{\rm TeV}}\right)^2 + 10.38 + 1.44 \left(\frac{s_{L}^{t}}{0.1}\right)^{2} \left(\frac{ m_T}{{\rm TeV}}\right)^2 \right \}.
\end{eqnarray}
In Fig.~\ref{fig:mw7}, the predictions of the $W$-boson mass, as well as various limitations, are presented. It is shown that although the theory can explain the mass anomaly in the parameter space, consistent with the constraints from the oblique parameters, the $Z b \bar{b}$ constraints prefer a significantly smaller $s_L^t$ for $m_T \lesssim 5~{\rm TeV}$ in the standard average scenario. As a result, a feasible solution to the anomaly sets a clear lower bound on $m_T$, i.e., $m_T \gtrsim 7~{\rm TeV}$ ($m_T \gtrsim 6~{\rm TeV}$) if the theory is required to explain the mass anomaly at $1\sigma$ ($2\sigma$) level. Evidently, this value is much larger than the LHC bound.

\subsection{Other issues of the VLQ extensions}

Up to present, there are many studies of the oblique parameters in VLQ extensions (see, e.g., Refs.~\cite{Lavoura:1992np,Aguilar-Saavedra:2013qpa,Arhrib:2016rlj,Chen:2017hak,He:2022zjz}) and some of them, however, are discrepant. So it is essential to clarify the subtleties in the involved calculations. The oblique parameters were first computed in the framework of the singlet and doublet VLQ extensions in Ref.~\cite{Lavoura:1992np}, and their expressions were universally formulated  in terms of generalized Cabibbo-Kobayashi-Maskawa mixing matrices. As far as the doublet VLQ extensions are concerned, these expressions were proved to be correct through numerical calculations by both us and A. Arhrib, the first author of Ref.~\cite{Arhrib:2016rlj}. They were also reproduced analytically in Ref.~\cite{He:2022zjz}. In particular, the author of Ref.~\cite{He:2022zjz} pointed out that the derivation of the universal expressions relies on the relations in Eqs. (7-10) of Ref.~\cite{Lavoura:1992np}. This observation implies that the expressions, especially those for $S$ and $U$ parameters, cannot be applied to the triplet extensions $T_1$ and $T_2$ since the relations are not satisfied in these extensions. Following this logic, the results of Ref.~\cite{Chen:2017hak}, where the oblique parameters for the triplet extensions were calculated by the formula in Ref.~\cite{Lavoura:1992np}, may be problematic\footnote{In fact, there is another evident mistake in Ref.~\cite{Chen:2017hak}, i.e., the invalid relations in its Eq.~(A.20), $m_X^2 = (c_R^b)^2 m_B^2 + (s_R^b)^2 m_b^2 = (c_R^t)^2 m_T^2 + (s_R^t)^2 m_t^2 $, were used in calculation.}. This point was recently verified by A. Arhrib. Besides these details, it should be noted that the $T_2$ extension was also considered to explain the mass anomaly~\cite{He:2022zjz}. Although that work obtained independently the same formula for the $S$ and $T$ parameters as ours, its Fig. 3 showed that the extension could explain the anomaly even for $m_T < 3~{\rm TeV}$, which is opposite to our conclusions. The main reason for the difference is that the $Z b \bar{b}$ coupling constraints were not considered in limiting $\sin \theta_L^t$ in Ref.~\cite{He:2022zjz}. In addition, it neglected the contribution of the $U$ parameter to $m_W$, which is sizable by our calculation.

Concerning our approximations of the oblique parameters, the following observations are in order:
\begin{itemize}
\item They take into account all the dominant contributions of the exact formulas, which are calculated by the loop functions in Eq.~(\ref{fun}). Numerically, the induced difference for all the quantities is found to be less than $10\%$ for
$s_{L,R}^t < 0.2 $ and $m_T > 1~{\rm TeV}$.
\item The formulas for the $T$ parameter agree with those in Ref.~\cite{Chen:2017hak}, except the $s_{L,R}^4 m_T^2$ terms are neglected for the $U$, $D$, $Q_1$, $Q_7$, $T_1$, and $T_2$ extensions in Ref.~\cite{Chen:2017hak}. Concerning the expression of the $T$ parameter for the $Q_5$ extension, there is a minor typo in Eq.~(3.9) of Ref.~\cite{Chen:2017hak}; namely, the factor $(s_R^b)^5$ should be replaced by $(s_R^b)^4$.
\item The formulas for the $S$ and $U$ parameters in the $U$, $D$, $Q_1$, $Q_5$, and $Q_7$ extensions agree with those in Ref.~\cite{Chen:2017hak} except for the $S$ parameter in the $Q_5$ and $Q_7$ extensions. This difference does not affect $\delta m_W$ essentially, as it is dominated by the $T$-term contribution.
\item This study cannot reproduce the formulas of Ref.~\cite{Chen:2017hak} for the $S$ and $U$ parameters in the $T_1$ and $T_2$ extensions due to the reasons discussed above.
\end{itemize}

Finally, the following related aspects of the VLQ extensions are commented on briefly.
\begin{itemize}
\item It would be pretty reasonable to expect that, if VLQs exist, then so could be vectorlike leptons (VLLs). In this case, the oblique parameters are also contributed to by the VLL loops. In the extension of the SM with only one VLL multiplet, the form of this additional contribution to $T$ parameter is similar to the results of this work. It has the following property~\cite{Lee:2022nqz}
    \begin{eqnarray}
    T_{\rm VLL} \propto \frac{m_L^2}{m_Z^2} \sin^4 \theta_L \propto \frac{\xi_L^2 v^2}{m_Z^2} \sin^2 \theta_L,
    \end{eqnarray}
    if terms proportional to SM lepton mass are neglected, where $m_L$ denotes the VLL mass and $\theta_L$ is the VLL-lepton mixing angle. Since the VLL-lepton Yukawa coupling $\xi_L$ is naturally much smaller than $\xi_j$ in Eq.~(\ref{Yukawa}) and $\sin \theta_L$ has been tightly limited by the EWPOs, $T_{\rm VLL}$ is negligibly small~\cite{Lee:2022nqz}. It is emphasized that this conclusion is valid only for the minimal VLL extension. To be specific, if the SM is extended by more than one multiplet, then the VLL contribution may get enhanced when the mixing and mass splitting between VLLs are large~\cite{Cynolter:2008ea}. This feature was recently utilized in Refs.~\cite{Kawamura:2022uft,Baek:2022agi,Zhou:2022cql} to explain the $W$-boson mass anomaly and the muon $g-2$ anomaly simultaneously.
\item As mentioned before, the introduction of the VLQs will exacerbate the vacuum stability problem of the SM when the running of the Higgs quartic coupling by the renormalization group equation (RGE) is considered. This problem can be solved or at least alleviated, by adding bosonic freedom into the theory. As far as the minimal VLQ extensions are concerned, the most economic way is using the singlet scalar field, which is responsible for the bare mass of VLQs, to change the RGE behavior of the coupling. It has been shown that this strategy works well for the $U$, $D$, $Q_1$, and $T_2$ extensions in broad parameter space, but it is inefficient for the $Q_7$ extension~\cite{Arsenault:2022xty}. The latter situation may be improved by invoking more complex frameworks.
\end{itemize}

\section{CONCLUSIONS}
\label{sec:split}
\label{sec:dec}

In many attractive BSM theories, VLQs are used to solve some fundamental problems in particle physics. They may couple directly with gauge bosons in the SM, and their mixings with the third-generation quarks can significantly affect EWPOs. Motivated by this feature, seven economic VLQ extensions that predict renormalizable $H \bar{Q} q$ interactions are considered to interpret the recently observed $W$-boson mass anomaly. For each extension, the $W$-boson mass are calculated, and the $S$, $T$, and $U$ parameters are used to limit the parameter space of the extensions. Other important constraints, such as those from the $Z b \bar{b}$ couplings on the $D$, $Q_5$, $T_1$, and $T_2$ extensions, are also implemented. The following conclusions are obtained:
\begin{itemize}
\item The $U$ and $Q_1$ extensions can interpret the anomaly in their broad parameter space. The explanations are consistent with the constraints from the oblique parameters at the $2\sigma$ level in the standard average scenario and at the $1\sigma$ level in the conservative average scenario. They also satisfy other experimental constraints, such as the measurements of the properties for the top quark, bottom quark, and Higgs boson, the LHC search for VLQs, and the perturbativity of the theories. The typical size of the involved Yukawa coupling is around 1, which is comparable to the top quark Yukawa coupling in the SM.
\item In the $T_2$ extension, the $b^\prime-B^\prime$ mixing is correlated with the $t^\prime-T^\prime$ mixing. Since the former mixing can significantly alter the $Z b_L \bar{b}_L$ coupling and is thus tightly limited by the coupling measurements, the $t^\prime-T^\prime$ mixing must be small. As a result, although the $T_2$ extension can explain the anomaly in the parameter space, consistent with the constraints from the oblique parameters, the $Z b \bar{b}$ couplings exclude the explanation when $m_T \lesssim 5~{\rm TeV}$. Fortunately, they become compatible if $m_T \gtrsim 6~{\rm TeV}$ in the standard average scenario, a value much larger than the LHC constraints on $m_T$.
\item Although the $Q_5$ extension can explain the mass anomaly in the parameter space, consistent with the constraints from the oblique parameters, it needs an unnaturally large coupling of the right-handed bottom quark to new physics. Thus, it is theoretically less appealing.
\item Owing to the assignment of the charges for the $SU(3)_C \times SU(2)_L \times U(1)_Y$ gauge groups, the $D$ extension predicts a negative correction to $W$-boson mass for small VLQ-quark mixing, so it cannot explain the anomaly after considering the $Z b \bar{b}$ coupling constraints. This situation is also applied to the $T_1$ extension.
\item Concerning the $Q_7$ extension, there remains very narrow parameter space that can explain the mass anomaly and at the same time, keep consistent with the constraints from the oblique parameters and the $Z b \bar{b}$ couplings. This parameter space, however, is characterized by an excessively large Yukawa coupling, $\xi_{Q_7} \sim 3$, which exacerbates greatly the vacuum stability problem of the SM.
\end{itemize}
It is emphasized that if the $W$-boson mass anomaly is confirmed by other experiments in the future, these obtained conclusions may serve as a useful guide for model building, which is the main merit of this study.

\section*{ACKNOWLEDGMENTS}
We thank Professors Abdesslam Arhrib, Yi Liao, and Jianghao Yu and Dr. Shi-Ping He for their helpful discussion. This work is supported by the National Natural Science Foundation of China (NNSFC) (Grants No. 11705048 and No. 12075076), the National Research Project Cultivation Foundation of Henan Normal University (Grants No. 2020PL16 and No. 2021PL10), and the High
Performance Computing Center of Henan Normal
University.

\section*{APPENDIX: W/Z COUPLINGS IN THE VLQ EXTENSIONS}

The SM Lagrangian for the W/Z couplings to the third-generation quarks is modified by their mixing with VLQs~\cite{Aguilar-Saavedra:2013qpa}. In the following, the third-generation-quark-dominated mass eigenstates are denoted as $t/b$, and the VLQ-dominated states are denoted as $Q/Q'=X,T,B,Y$. They are collectively called light quarks and heavy quarks, respectively, for brevity.

The renewed Lagrangian for light quarks takes the following form:
\begin{eqnarray}
\mathcal{L}_W & = & -\frac{g}{\sqrt 2} \bar t \gm \left( V_{tb}^L P_L + V_{tb}^R P_R \right) b \Wm^+ +\text{H.c.},  \notag \\
\mathcal{L}_Z & = & -\frac{g}{2 c_W} \bar t \gm \left( X_{tt}^L P_L + X_{tt}^R P_R \right) t \Zm -\frac{g}{2 c_W} \bar b \gm \left( X_{bb}^L P_L + X_{bb}^R P_R  \right) b \Zm \notag,
\label{ec:ll}
\end{eqnarray}
where the charged current form factors ($V$) and the neutral current form factors ($X$) are listed in Table~\ref{tab:llW} and~\ref{tab:ll}, respectively.

\begin{table}[t]
	\begin{center}
		\begin{tabular}{c|cc}
			& $V_{tb}^L$ & $V_{tb}^R$
			\\ \hline
			$U$ & $\clx$ & 0 \\
            $D$ & $c_L^b$ & 0 \\
			$Q_1$ & $\clu \cld + \slu \sld$ & $\sru \srd$
			\\
			$Q_5$ & $c_L^b$ & 0
			\\
			$Q_7$ & $\clx$ & 0
			\\
			$T_1$ & $\clu \cld + \sqt \slu \sld$ & $\sqt \sru \srd$
			\\
			$T_2$ & $\clu \cld + \sqt \slu \sld$ & $\sqt \sru \srd$
		\end{tabular}
		\caption{Light-light quark couplings to the $W$ boson.}
		\label{tab:llW}
	\end{center}
\end{table}

\begin{table}[t]
	\begin{center}
   \footnotesize
		\begin{tabular}{c|cccc}
			 & $X_{tt}^L$ & $X_{tt}^R$ & $X_{bb}^L$ & $X_{bb}^R$
			\\ \hline
			$U$ & $(\clx)^2 - 2 Q_t s_W^2 $ & $- 2 Q_t s_W^2 $  & $-1- 2 Q_b s_W^2$ & $- 2 Q_b s_W^2$
			\\
            $D$ & $1- 2 Q_t s_W^2 $ & $- 2 Q_t s_W^2$ & $- (c_L^b)^2 - 2 Q_b s_W^2$ & $- 2 Q_b s_W^2$ \\
			$Q_1$ & $1- 2 Q_t s_W^2 $ & $(\sru)^2- 2 Q_t s_W^2 $  & $-1- 2 Q_b s_W^2$ & $-(\srd)^2- 2 Q_b s_W^2$
			\\
            $Q_5$ & $1- 2 Q_t s_W^2 $ & $- 2 Q_t s_W^2$ & $- (c_L^b)^2 + (s_L^b)^2 - 2 Q_b s_W^2$ & $ (s_R^b)^2 - 2 Q_b s_W^2$ \\
			$Q_7$ & $(\clx)^2-(\slx)^2 - 2 Q_t s_W^2 $ & $-(\srx)^2 - 2 Q_t s_W^2 $   & $-1- 2 Q_b s_W^2$ & $- 2 Q_b s_W^2$
			\\
			$T_1$  & $1+(\slu)^2- 2 Q_t s_W^2 $ & $2(\sru)^2 - 2 Q_t s_W^2$  & $-(\cld)^2- 2 Q_b s_W^2$ & $- 2 Q_b s_W^2$ \\
			$T_2$  & $(\clu)^2- 2 Q_t s_W^2 $ & $- 2 Q_t s_W^2$  & $-1-(\sld)^2- 2 Q_b s_W^2$ & $-2(\srd)^2- 2 Q_b s_W^2$
		\end{tabular}
		\caption{Light-light quark couplings to the $Z$ boson.}
		\label{tab:ll}
	\end{center}
\end{table}

Similarly, the Lagrangian for the heavy states, $Q,Q'=X,T,B,Y$, is given by
\begin{eqnarray}
\mathcal{L}_W & = & -\frac{g}{\sqrt 2} \bar Q \gm \left( V_{QQ'}^L P_L + V_{QQ'}^R P_R \right) Q' \Wm^+ +\text{H.c.} \notag, \\
\mathcal{L}_Z & = & -\frac{g}{2 c_W} \bar Q \gm \left(  X_{QQ}^L P_L + X_{QQ}^R P_R  \right) Q \Zm  \notag,
\label{ec:HH}
\end{eqnarray}
where the charged current form factors ($V$) are presented in Table~\ref{tab:hhW}, and the neutral current form factors ($X$) are given in Tables~\ref{tab:XXZ},~\ref{tab:TTZ}, and~\ref{tab:BBZ}.
\begin{table}[t]
	\begin{center}
		\begin{tabular}{c|cccccc}
			& $V_{XT}^L$ & $V_{XT}^R$ & $V_{TB}^L$ & $V_{TB}^R$  & $V_{BY}^L$ & $V_{BY}^R$
			\\ \hline
			$Q_1$ & $\cdots$ & $\cdots$ & $\clu \cld + \slu \sld $ & $\cru \crd$ & $\cdots$ & $\cdots$
			\\
			$Q_5$ & $\cdots$ & $\cdots$ & $\cdots$ & $\cdots$ & $c_L^b$ & $c_R^b$
			\\
			$Q_7$ & $\clx$ & $\crx$ & $\cdots$ & $\cdots$ & $\cdots$ & $\cdots$
			\\
			$T_1$ & $\cdots$ & $\cdots$ & $\slu \sld + \sqt \clu \cld$ & $\sqt \cru \crd$ & $\sqt \cld$ & $\sqt \crd$ \\
			$T_2$ & $\sqt \clu$ & $\sqt \cru$ & $\slu \sld + \sqt \clu \cld$ & $\sqt \cru \crd$ & $\cdots$ & $\cdots$
		\end{tabular}
		\caption{Heavy-heavy quark couplings to the $W$ boson.}
		\label{tab:hhW}
	\end{center}
\end{table}

\begin{table}[t]
	\begin{center}
		\begin{tabular}{c|cc}
			& $X_{XX}^L$ & $X_{XX}^R$
			\\ \hline
			$Q_7$ &$ 1- 2 Q_X s_W^2$ & $1- 2 Q_X s_W^2$
			\\
			$T_2$ & $2- 2 Q_X s_W^2$ & $2- 2 Q_X s_W^2$
		\end{tabular}
		\caption{$ZXX$ couplings in the VLQ extensions. }
		\label{tab:XXZ}
	\end{center}
\end{table}

\begin{table}[t]
	\begin{center}
		\begin{tabular}{c|cc}
			& $X_{TT}^L$ & $X_{TT}^R$
			\\ \hline
			$U$ & $(\slx)^2- 2 Q_T s_W^2$ & $- 2 Q_T s_W^2$
			\\
			$Q_1$ & $1- 2 Q_T s_W^2$ & $(\cru)^2- 2 Q_T s_W^2$
			\\
			$Q_7$  & $(\slx)^2 - (\clx)^2- 2 Q_T s_W^2$ & $-(\crx)^2- 2 Q_T s_W^2$
			\\
			$T_1$  & $1+(\clu)^2- 2 Q_T s_W^2$ & $2(\cru)^2 - 2 Q_T s_W^2$ \\
			$T_2$  & $(\slu)^2- 2 Q_T s_W^2$ & $- 2 Q_T s_W^2$
		\end{tabular}
		\caption{$ZTT$ couplings in the VLQ extensions.}
		\label{tab:TTZ}
	\end{center}
\end{table}

\begin{table}[t]
	\begin{center}
\footnotesize
		\begin{tabular}{c|cccc}
			& $X_{BB}^L$ & $X_{BB}^R$  & $X_{YY}^L$ & $X_{YY}^R$
			\\ \hline
            $D$ & $-(s_L^b)^2 - 2 Q_B s_W^2$ & $- 2 Q_B s_W^2$  & $\cdots$ & $\cdots$
			\\
			$Q_1$  & $-1- 2 Q_B s_W^2$ & $-(\crd)^2- 2 Q_B s_W^2$   & $\cdots$ & $\cdots$
			\\
			$Q_5$  & $-(\sld)^2 + (c_L^b)^2 - 2 Q_B s_W^2$ & $ (c_R^b)^2 - 2 Q_B s_W^2$   & $- 1 - 2 Q_Y s_W^2$ & $-1 -2 Q_Y s_W^2$ \\
			$T_1$  & $-(\sld)^2- 2 Q_B s_W^2$ & $- 2 Q_B s_W^2$   & $-2-2 Q_Y s_W^2$ & $-2-2 Q_Y s_W^2$ \\
			$T_2$  & $-1-(\cld)^2- 2 Q_B s_W^2$ & $-2 (\crd)^2- 2 Q_B s_W^2$   & $\cdots$ & $\cdots$
		\end{tabular}
		\caption{$ZBB$ and $ZYY$ couplings in the VLQ extensions.}
		\label{tab:BBZ}
	\end{center}
\end{table}

\begin{table}[t]
	\begin{center}
		\begin{tabular}{c|cccc}
			& $V_{Xt}^L$ & $V_{Xt}^R$ & $V_{Tb}^L$ & $V_{Tb}^R$
			\\ \hline
			$U$ & $\cdots$ & $\cdots$ & $\slx $ & 0
			\\
			$Q_1$ & $\cdots$ & $\cdots$ & $\slu \cld  - \clu \sld $ & $-\cru \srd $
			\\
			$Q_7$ & $-\slx $ & $-\srx $ & $\slx $ & 0
			\\
			$T_1$ & $\cdots$ & $\cdots$ & $\slu \cld - \sqt \clu \sld$ & $-\sqt \cru \srd $ \\
			$T_2$ & $-\sqt \slu $ & $-\sqt \sru $ & $\slu \cld - \sqt \clu \sld$ & $-\sqt \cru \srd $
		\end{tabular}
		\caption{Charged current for heavy-light quark transition in VLQ extensions.}
		\label{tab:WHl}
	\end{center}
\end{table}

\begin{table}[t]
	\begin{center}
    \footnotesize
		\begin{tabular}{c|cccc|cccc}
			&\multicolumn{4}{c|}{Heavy-light coupling to $W$ boson.}&\multicolumn{4}{c}{Heavy-light coupling to $Z$ boson.}
			\\ \cline{2-9}
			& $V_{tB}^L$ & $V_{tB}^R$  & $V_{bY}^L$ & $V_{bY}^R$   & $X_{tT}^L$ & $X_{tT}^R$ & $X_{bB}^L$ & $X_{bB}^R$
			\\ \hline
			$U$ &  $\cdots$   &  $\cdots$   &  $\cdots$   &  $\cdots$ & $\slx \clx $ & 0 & $\cdots$ & $\cdots$
			\\
            $D$ &  $s_L^b$   &  0   &  $\cdots$   &  $\cdots$ & $\cdots$ & $\cdots$ & $-s_L^b c_L^b$ & 0
			\\
			$Q_1$ & $\clu \sld -\slu \cld $ & $-\sru \crd$   &  $\cdots$   &  $\cdots$ & 0 & $-\sru \cru $ & 0 & $\srd \crd $
			\\
            $Q_5$ &  $s_L^b$   &  0   &  $-s_L^b$   & $-s_R^b$ & $\cdots$ & $\cdots$ & $-2 s_L^b c_L^b$ & $-s_R^b c_R^b $
			\\
			$Q_7$ & $\cdots$ &  $\cdots$   &  $\cdots$   &  $\cdots$  & $2 \slx \clx $ & $\srx \crx $ & $\cdots$ & $\cdots$
			\\
			$T_1$ & $\clu \sld - \sqt \slu \cld$ & $-\sqt \sru \crd $  &  $-\sqt \sld$   &  $-\sqt \srd$  & $-\slu \clu $ & $-2\sru\cru$ & $-\sld \cld $ & 0 \\
			$T_2$ & $\clu \sld - \sqt \slu \cld$   & $-\sqt \sru \crd $ &  $\cdots$   &  $\cdots$  & $\slu \clu $ & 0 & $\sld \cld $ & $2\srd \crd$
		\end{tabular}
		\caption{Neutral current for heavy-light quark transitions in VLQ extensions.}
		\label{tab:lH}
	\end{center}
\end{table}

Finally, the Lagrangian involving heavy-light quark transition is
\begin{eqnarray}
\mathcal{L}_W & = & -\frac{g}{\sqrt 2} \bar Q \gm \left( V_{Qq}^L P_L + V_{Qq}^R P_R \right) q \Wm^+ +\text{H.c.} \notag \\
& & -\frac{g}{\sqrt 2} \bar q \gm \left( V_{qQ}^L P_L + V_{Qq}^R P_R \right) Q \Wm^+ +\text{H.c.}  \notag, \\
\mathcal{L}_Z & = & -\frac{g}{2 c_W} \bar q \gm \left(  X_{qQ}^L P_L + X_{qQ}^R P_R \right) Q \Zm +\text{H.c.}  \notag,
\end{eqnarray}
where the factors $V$ and $X$ are listed in Table~\ref{tab:WHl} and Table~\ref{tab:lH}, respectively.

\end{document}